\documentclass[aps,pra,superscriptaddress,twocolumn,amsmath,amssymb]{revtex4-2}
\usepackage{graphicx}
\usepackage{amsmath,amssymb}
\usepackage{dcolumn}
\usepackage{mathrsfs}
\usepackage{physics}
\usepackage{float}
\usepackage{booktabs}
\usepackage{makecell}
\usepackage{hyperref}
\usepackage{color}
\newcommand{\Lb}[2]{\mathcal{L}_{#1#2}}

\graphicspath{{figures/}}

\begin{document}

\title{Superradiant photon bunching in nanolasers: from a polarization-order hierarchy of the master equation to Dicke Markov chains}
\date{\today}

\author{Matias Bundgaard-Nielsen}
\affiliation{Department of Electrical and Photonics Engineering, Technical University of Denmark, Building 343, 2800 Kongens Lyngby, Denmark}
\affiliation{NanoPhoton - Center for Nanophotonics, Technical University of Denmark, Building 343, 2800 Kongens Lyngby, Denmark}
\author{Jesper M\o rk}
\email[]{jesm@dtu.dk}
\affiliation{Department of Electrical and Photonics Engineering, Technical University of Denmark, Building 343, 2800 Kongens Lyngby, Denmark}
\affiliation{NanoPhoton - Center for Nanophotonics, Technical University of Denmark, Building 343, 2800 Kongens Lyngby, Denmark}

\begin{abstract}
In nanolasers, a few to a few hundred quantum emitters share a single mode of light. Coupled identically to the mode, the emitters do not radiate independently but form collective bright and dark states, producing superradiant bursts and strongly bunched photon statistics below the laser threshold. We investigate to what extent this behavior can be understood as random jumps between collective Dicke states. From the laser master equation, we derive a hierarchy of models defined by the polarization order, i.e., the number of photons involved in the coherence between the emitters and the light, and include the coherence to various orders. With all of the coherences eliminated, we obtain a Dicke Markov chain, a stochastic jump process over the photon number and the Dicke states, in which collective effects enter only through enhanced or suppressed transition rates. It predicts superthermal bunching and a strong even--odd dependence on the emitter number. It recovers about 80\% of the bunching in the most coherent regime, where conventional rate equations predict no superthermal bunching. Keeping the first polarization order, i.e., the optical polarization, and eliminating the second order, i.e., the two-photon coherences, yields a population-polarization model, which recovers the bunching of the full master equation almost perfectly. This hierarchy of polarization orders is an alternative to the cumulant expansion and, unlike it, recovers the master equation at the highest order. As a numerically exact reference, a symmetry-reduced master equation extends the simulation range all the way to 34 (24) emitters below (above) threshold. The Dicke Markov chain can be solved directly for hundreds of emitters and sampled for thousands of emitters, further extending the range of numerical methods. We compare the numerical performance of the methods in detail to guide the choice of technique.
\end{abstract}
\maketitle

\section{Introduction}
Nanolasers confine light to volumes where a few, or a few hundred, quantum emitters interact with a single cavity mode \cite{Ma2019ApplicationsNanolasers,Dimopoulos2022Electrically-DrivenThreshold}. In this regime, a large fraction of the spontaneous emission is funneled into the lasing mode and the transition to lasing extends over a broad range of pump rates \cite{Rice1994PhotonAnalogy,Saldutti2024TheNanolasers}, where the photon statistics are strongly influenced by quantum noise \cite{Mork2020SqueezingConfinement}. In addition, when several emitters couple identically to the cavity, they no longer emit independently: the ensemble develops collective bright and dark states, first analyzed by Dicke \cite{Dicke1954CoherenceProcesses,Gross1982SuperradianceEmission}, leading to superradiant enhancement or subradiant suppression of the emission. Signatures of such collective effects have been predicted and observed in quantum-dot nanolasers, most strikingly as superthermal photon bunching, $g^{(2)}(0)\gg2$, below threshold \cite{Temnov2005SuperradianceCavity,Auffeves2011FewIndividualization,Leymann2015Sub-Nanolasers,Jahnke2016GiantNanolasers,Kreinberg2017EmissionCoupling,Zhang2018MonteLasing, Andre2019OnNano-LEDs,Zhang2021UltranarrowStates,Protsenko2021QuantumNanolasers, Vyshnevyy2022Gain-dependentNanolasers}, and superradiance is the operating principle of proposed ultranarrow-linewidth lasers based on cold atoms \cite{Meiser2009ProspectsLaser,Bohnet2012ALaser,Norcia2016SuperradianceLaser,Debnath2018LasingRegime}.

Simulating these systems is demanding. The Hilbert space of a quantum master equation (ME) grows exponentially with the number of emitters, and even permutationally invariant (PI) formulations \cite{Chase2008CollectiveParticles,Baragiola2010CollectiveSystems,Gegg2016EfficientSystems,Kirton2017SuppressingDecay,Shammah2018OpenInvariance}, which reduce the scaling to polynomial, become impractical beyond a few tens of emitters once the photon numbers grow at and above the laser threshold. Cumulant expansions \cite{Gies2007SemiconductorLasers,Plankensteiner2022quantumcumulantsjl} reach larger systems but require a careful choice of truncation order and give limited access to full counting statistics. Semiclassical Langevin rate equations \cite{Coldren1997} fail in the few-photon regime where nanolasers operate \cite{Bundgaard-Nielsen2023StochasticNanolaser,Mork2018RateEmitters}.

\begin{figure*}[!ht]
    \centering
    \includegraphics[width=0.95\linewidth]{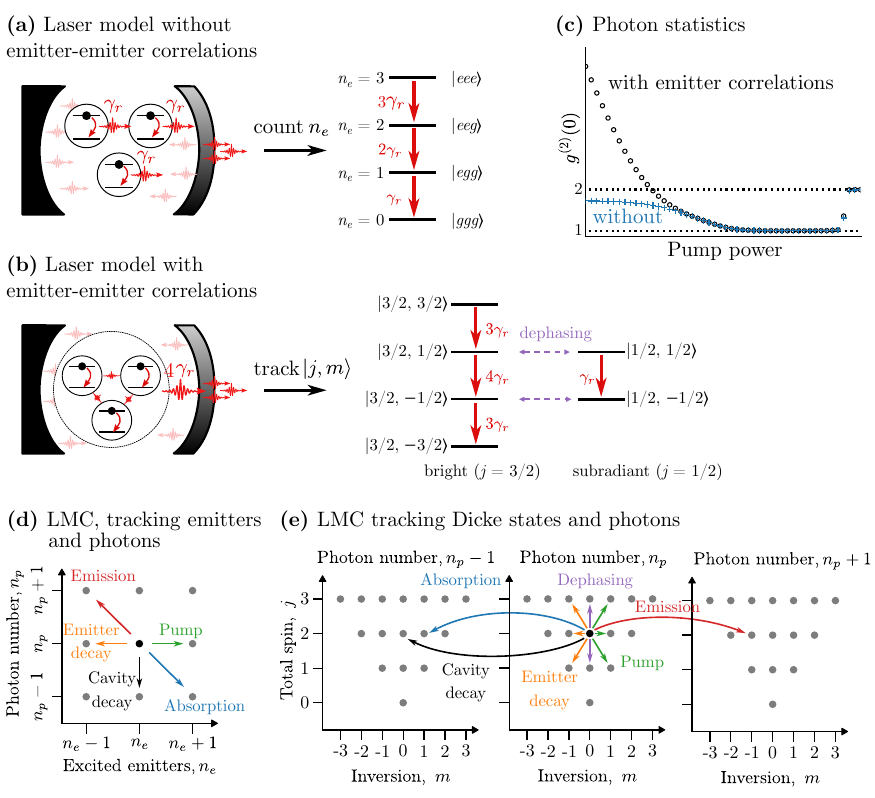}
    \caption{(a) Schematic of a laser model without emitter--emitter correlations. (b) Schematic of a laser model with emitter--emitter correlations and thus superradiance. (c) Typical second-order autocorrelation function $g^{(2)}(0)$ for a laser model with and without superradiance. Below threshold, superradiance leads to $g^{(2)}(0)\gg2$, i.e., strong bunching of the emitted photons. (d) State space and events of the conventional Laser Markov Chain (LMC), which tracks the photon number $n_p$ and the number of excited emitters $n_e$. Arrows show the birth--death events of Table~\ref{tab:events_dicke} with the uncorrelated-emitter rates. (e) State space and events of the Dicke LMC, which replaces the single emitter variable $n_e$ by the pair $(j,m)$ of collective spin and inversion, so that the full state space consists of one Dicke triangle (shown for $n_0=6$ emitters) per photon number $n_p$. Local pump (green), decay (orange), and dephasing (purple) events act within a photon slice and transfer the ensemble between neighboring $j$ ladders; collective emission and absorption (red/blue) carry the Dicke enhancement factors $C_\pm(j,m)$ and connect neighboring slices, as does cavity decay (black). The states with small $j$ are subradiant; the $j=0$ states are fully dark.}
    \label{fig:schematic}
\end{figure*}

An attractive middle ground is provided by stochastic rate equations, where the laser rate equations are interpreted as a birth--death process for the integer-valued photon and excited-emitter populations \cite{Rice1994PhotonAnalogy,Puccioni2015StochasticLasing,Mork2018RateEmitters,Andre2020}. We refer to this description as the Laser Markov Chain (LMC) \cite{Bundgaard-Nielsen2026ModelingChains}. The LMC follows rigorously from the ME by keeping only the diagonal of the density matrix and adiabatically eliminating the optical polarization \cite{Bundgaard-Nielsen2026ModelingChains}, and it reproduces master-equation photon statistics of few-emitter lasers remarkably well \cite{Bundgaard-Nielsen2023StochasticNanolaser}, including, with a simple extension, the laser linewidth \cite{Bundgaard-Nielsen2025SimpleNanolasers}. Its computational cost is essentially independent of the number of emitters, and in the many-particle limit, it reduces to the familiar Langevin rate equations \cite{Bundgaard-Nielsen2026ModelingChains}. However, the LMC state $(n_p, n_e)$ has no information about emitter--emitter correlations: all microstates with the same number of excited emitters are treated as equivalent, and collective effects are absent by construction.

This raises the question that motivates the present work: to what extent can superradiant photon bunching be understood as random jumps between collective emitter states, with modified rates but without coherence between the emitters and the light? The Dicke states are themselves quantum mechanical: they are entangled states of the emitters, and their brightness reflects the correlations between them. Once the ensemble is described by the occupation probabilities of these states, however, superradiant bursts and dark-state trapping acquire a simple classical stochastic interpretation, as random switching between bright and dark states with collectively enhanced or suppressed emission rates. By extending the LMC to keep track of these states, we show that this simple stochastic picture is remarkably accurate when compared with an exact master equation.

We address this question by eliminating the coherence between the emitters and the light field from the master equation step by step. We expand the master equation in the collective Dicke states and sort the density matrix by its polarization order, i.e., the number of photons involved in the coherence. Keeping all polarization orders gives a numerically exact, symmetry-reduced master equation, which we solve for up to 34 (24) emitters below (above) threshold, compared with 10 (5) using previous symmetry-reduced approaches \cite{Shammah2018OpenInvariance} for the same problem. At the opposite end, we find that eliminating all coherences yields a Markov chain over the states $(n_p, j, m)$, the Dicke LMC, in which $j$ tracks the collective brightness of the gain medium and $m$ the inversion (see Fig.~\ref{fig:schematic}). In between, keeping the optical polarization as an explicit variable while eliminating the two-photon coherences gives what we call the population--polarization model.

The Dicke LMC reproduces the superradiant physics that conventional rate equations miss: superthermal photon bunching \cite{Jahnke2016GiantNanolasers} and a strong even--odd dependence on the emitter number in the photon bunching \cite{Auffeves2011FewIndividualization}. Below threshold, without any dephasing or background decay, it recovers about $80\%$ of the bunching, and it becomes accurate once dephasing exceeds the cavity loss rate. The population--polarization model recovers most of the remainder, reducing the error in $g^{(2)}(0)$ significantly at an order of magnitude lower cost than the cost of the master equation. This correction, however, cannot be formulated into a Markov chain. Most of the superradiant bunching can thus be understood as random switching between bright and dark collective states with modified rates, whereas the remainder is an interference effect between successive emissions that requires coherence between the emitters and the light.

We also compare with the cumulant expansion of QuantumCumulants.jl~\cite{Plankensteiner2022quantumcumulantsjl}, a widely used approach for many-emitter systems. Below threshold, the fourth-order expansion, the lowest order that can describe bunching, overestimates $g^{(2)}(0)$ by about $50\%$ for ten emitters and misses the even--odd effect, although it is accurate far above threshold and for strong dephasing. Finally, we benchmark the accuracy and speed of the available methods for many-emitter lasers.

The paper is organized as follows. In Sec.~\ref{sec:model}, we introduce the laser master equation, its symmetry-reduced representation in terms of collective Dicke states, and the polarization orders used to derive the Dicke LMC and the population--polarization model. In Sec.~\ref{sec:elimination}, we derive the two models by adiabatically eliminating the first and second polarization order, respectively. In Sec.~\ref{sec:results}, we benchmark the models across the transition to lasing, as a function of dephasing and emitter number, for delayed correlations, and in terms of computational cost. Finally, in Sec.~\ref{sec:conclusion}, we conclude and provide an outlook.

Table~\ref{tab:methods} collects all the methods compared in this work, with their state spaces, solution methods, demonstrated system sizes, and accuracies, and the reader may find it useful as a guide to the naming used in the figures. The code for all methods in Table~\ref{tab:methods}, together with the data and scripts that reproduce the figures of this paper, is available online~\cite{Bundgaard-Nielsen2026SuperradiantLMCCode}.

\begin{table*}[!t]
\caption{Overview of the methods compared in this work. All methods solve the same laser master equation, Eq.~\eqref{eq:me}, but they differ in the approximations (if any) made. ``Solution'' distinguishes direct stationary solves of the linear dynamical equation ($\dot x = M x = 0$), trajectory sampling with Gillespie's first reaction method (GFRM), and time integration to the steady state, see Appendix~\ref{app:numerical_details}. ``Largest $n_0$'' is the largest emitter number for which a converged steady state was obtained in this work, below threshold ($\gamma_P = 10^{-4}\,\mathrm{ps}^{-1}$) and above threshold ($\gamma_P = 2\,\mathrm{ps}^{-1}$), with $\gamma_A = \gamma_D = 0$; the corresponding wall-clock times are given in Sec.~\ref{sec:results_cost}. For the direct solvers, the limit is set by memory and wall time, and it is lower above threshold, where the photon cutoff grows to thousands. ``Any'' marks methods whose cost does not grow with $n_0$. The error column refers to $g^{(2)}(0)$ in the collective regime below threshold, where the methods differ most; above threshold, all methods agree at the percent level.}
\label{tab:methods}
\footnotesize
\begin{ruledtabular}
\begin{tabular}{lllll}
\textbf{Method} & \textbf{State space} & \textbf{Solution} & \makecell[l]{\textbf{Largest $n_0$}\\ \textbf{(below/above)}} & \makecell[l]{\textbf{Error in}\\ $g^{(2)}(0)$} \\
\colrule
ME, PIQS \cite{Shammah2018OpenInvariance} & full Dicke + photon density matrix & direct & $10$ / $5$ & exact \\
ME, symmetry-reduced (Sec.~\ref{sec:model}) & $c=0$ sector, all polarization orders & direct & $34$ / $24$ & exact \\
Population--polarization model (Sec.~\ref{sec:secondorder}) & populations + polarization & direct & $\approx 100$ / $41$ & $\lesssim 1\%$ \\
Dicke LMC (Sec.~\ref{sec:lmc}) & $(n_p, j, m)$, classical & direct & $500$ / $60$ & $\sim 20\%$ \\
 & & GFRM & any / any & $\sim 20\%$ \\
Conventional LMC \cite{Mork2018RateEmitters,Bundgaard-Nielsen2026ModelingChains} & $(n_p, n_e)$, classical & direct & $\gg 1000$ / $1256$ & misses collectivity \\
Cumulant expansion, 4th order \cite{Plankensteiner2022quantumcumulantsjl} & 18 moments up to 4th order & time integration & any / any & \makecell[l]{$\sim50\%$} \\
\end{tabular}
\end{ruledtabular}
\end{table*}

\section{The laser master equation in the symmetry-reduced representation}
\label{sec:model}

For our superradiant laser model, we consider $n_0$ identical two-level emitters resonantly coupled to a single cavity mode. In a rotating frame, the coherent dynamics are governed by the Tavis--Cummings Hamiltonian \cite{TavisExactHamiltonian1968,Moelbjerg2013DynamicalEmitters}
\begin{equation}
    H = \sum_i g_i (a^\dagger \sigma_i + a \sigma_i^\dagger) = g\left(a^\dagger J_- + a J_+\right),
    \label{eq:hamiltonian}
\end{equation}
with $a$ being the photon annihilation operator, $g$ the light-matter coupling rate, and the collective spin operators $J_+ = \sum_i \sigma_i^{\dagger}$ and $J_- = \sum_i \sigma_i$, where $\sigma_i = \ket{g}_i\bra{e}_i$ is the transition operator of the $i$th emitter. Dissipation is included through the Lindblad master equation
\begin{equation}
\begin{aligned}
    \frac{\partial \rho}{\partial t} = \mathcal{L}\rho = -i[H,\rho] &+ \gamma_c \mathcal{D}_a[\rho] + \sum_i \gamma_A \mathcal{D}_{\sigma_i}[\rho] \\
    &+ \sum_i \gamma_P \mathcal{D}_{\sigma_i^\dagger}[\rho] + \sum_i \gamma_D \mathcal{D}_{\sigma_i^\dagger \sigma_i}[\rho],
    \label{eq:me}
\end{aligned}
\end{equation}
where $\mathcal{D}_A[\rho] = A\rho A^\dagger - \tfrac{1}{2}\{A^\dagger A, \rho\}$. Here $\gamma_c$ is the cavity decay rate, and $\gamma_A$, $\gamma_P$, and $\gamma_D$ denote the background decay, incoherent pump, and pure dephasing rates per emitter \cite{Mu1992One-atomLasers,Loffler1997SpectralLaser,Moelbjerg2013DynamicalEmitters}. We write $\mathcal{L} = \mathcal{L}_H + \mathcal{L}_D$, with $\mathcal{L}_H\rho = -i[H,\rho]$ and $\mathcal{L}_D$ collecting the dissipators.

\subsection{Collective Dicke states}
\label{sec:dicke}

Written in terms of the individual emitters, the emitter Hilbert space of Eq.~\eqref{eq:me} has dimension $2^{n_0}$, which, together with the cavity photon degrees of freedom, is intractable already for a few tens of emitters. The Hamiltonian in Eq.~\eqref{eq:hamiltonian} does not, however, address the emitters individually: because all $g_i = g$ are equal, the emitters enter only through the collective lowering operator $J_- = \sum_i \sigma_i$. Together with $J_+ = J_-^\dagger$ and the inversion operator $J_z = \tfrac{1}{2}\sum_i (\sigma_i^\dagger \sigma_i - \sigma_i \sigma_i^\dagger)$, these obey the angular-momentum algebra $[J_z, J_\pm] = \pm J_\pm$ and $[J_+, J_-] = 2J_z$, so the ensemble can be treated as a single large spin rather than as $n_0$ small ones.

The natural basis is then the joint eigenbasis of the commuting operators $J^2 = J_z^2 + \tfrac{1}{2}(J_+J_- + J_-J_+)$ and $J_z$, the Dicke states \cite{Dicke1954CoherenceProcesses,Gross1982SuperradianceEmission,Chase2008CollectiveParticles,Shammah2018OpenInvariance},
\begin{equation}
    J^2 \ket{j,m} = j(j+1)\ket{j,m}, \qquad
    J_z \ket{j,m} = m \ket{j,m} .
    \label{eq:dicke_def}
\end{equation}
Coupling $n_0$ spin-$1/2$ particles (two-level systems) produces total spins $j = n_0/2,\, n_0/2-1,\, \dots,\, j_{\min}$, with $j_{\min} = 0$ ($1/2$) for even (odd) $n_0$, and for each $j$ an inversion $m = -j, -j+1, \dots, j$. For $j<n_0/2$, each pair $(j,m)$ occurs $d^{n_0}_j$ times, with
\begin{equation}
    d^{n_0}_j = \binom{n_0}{n_0/2 - j}\frac{2j+1}{n_0/2+j+1},
    \label{eq:degeneracy}
\end{equation}
corresponding to the different number of ways of coupling the individual spins to the same total $j$. These copies do not carry additional information ($j$ and $m$ fully characterize them), but their presence needs to be taken into account when rates between states are calculated \cite{Chase2008CollectiveParticles,Shammah2018OpenInvariance}.

The two quantum numbers $m$ and $j$ have distinct physical meanings. The inversion $m$ simply counts excitations, $n_e = m + n_0/2$, and is the emitter variable that the conventional rate-equation description uses. The total spin $j$, in contrast, measures how coherently the excitations are shared among the emitters, and it has no classical counterpart. The distinction is already visible for $n_0 = 2$, where the four product states reorganize into a $j=1$ triplet and a $j=0$ singlet,
\begin{equation}
\begin{aligned}
    j=1:&\quad \ket{ee},\ \tfrac{1}{\sqrt2}\left(\ket{eg}+\ket{ge}\right),\ \ket{gg} , \\
    j=0:&\quad \tfrac{1}{\sqrt2}\left(\ket{eg}-\ket{ge}\right) .
\end{aligned}
    \label{eq:two_emitters}
\end{equation}
The two states with one excitation, $n_e = 1$, carry identical emitter populations and identical energy, yet they behave oppositely: the symmetric triplet member $\tfrac{1}{\sqrt2}\left(\ket{eg}+\ket{ge}\right)$ radiates into the cavity at twice the single-emitter rate, whereas the antisymmetric singlet $\tfrac{1}{\sqrt2}\left(\ket{eg}-\ket{ge}\right)$ satisfies $J_-\ket{j{=}0,m{=}0} = 0$ and cannot emit at all. It is thus a dark state.

In general, the collective operators act as ladder operators within a fixed $j$ ladder, $J_\pm \ket{j,m} = \sqrt{C_\pm(j,m)}\,\ket{j,m\pm1}$, with
\begin{equation}
    C_\pm(j,m) = (j \mp m)(j \pm m + 1) .
    \label{eq:cpm}
\end{equation}
Since emission into the cavity is governed by $J_-$, the rate at which a Dicke state radiates is proportional to the brightness $C_-(j,m) = (j+m)(j-m+1)$, which replaces the independent-emitter factor $n_e$ of the classical rate equations. For maximal spin, $j = n_0/2$, and half inversion, $m \approx 0$, one finds $C_- \approx n_0(n_0+2)/4$, which grows quadratically with the emitter number and exceeds the independent-emitter value $n_e = n_0/2$, which grows only linearly. The emission rate is thus enhanced, and this is superradiance. On the other hand, the $j = 0$ states hold $n_0/2$ excited emitters but have $C_- = 0$. They thus do not radiate at all but remain "dark". Correspondingly, all states with small $j$ are subradiant. 

This information on the degree of coherence among emitter states is not included in rate equations governing only the total photon number, $n_p$, and the total excited-emitter number, $n_e$. At fixed $n_e$, that is, at fixed $m$, the brightness $C_-(j,m)$ ranges from $0$ to $\sim n_0^2/4$ as $j$ varies over the allowed ladders, so no emission rate that depends on $n_e$ alone can represent it. Tracking the quantum number $j$ is thus a way to include super- and subradiance, and this is what the Dicke LMC of Sec.~\ref{sec:lmc} does.

\subsection{Reduction of state space}

Beyond expanding in the Dicke states, we can further reduce the state space of the master equation in Eq.~\eqref{eq:me} by noting that
\begin{equation}
    \rho_{n_p j m;\, n_p' j m'} \ne 0 \quad\text{only if}\quad n_p + m = n_p' + m' ,
    \label{eq:u1}
\end{equation}
which follows from the fact that the superoperator $\mathcal{C}[\rho] = N \rho - \rho N$ with $N = a^\dagger a + J_z$ commutes with the total Liouvillian, $\comm{\mathcal{C}}{\mathcal{L}} = 0$ (see Appendix \ref{app:symmetry}). Since $\mathcal{C}[\rho_{n_p j m;\, n_p' j m'}] = (n_p + m - (n_p' + m'))\, \rho_{n_p j m;\, n_p' j m'}$, each matrix element $\rho_{n_p j m;\, n_p' j m'}$ is an eigenoperator of $\mathcal{C}$ with eigenvalue $c = n_p + m - (n_p' + m')$, and $\comm{\mathcal{C}}{\mathcal{L}} = 0$ implies that $\mathcal{L}$ cannot mix eigenoperators with different $c$. Thus, if the system starts in the sector $c = 0$, i.e., $n_p + m = n_p' + m'$, which contains all diagonal states, the subsequent evolution remains confined to this sector. This not only reduces the state space needed to represent the density matrix $\rho$, but also gives a physical picture of the dynamics in terms of the polarization orders introduced below. A similar reduction is used for the emitters in PIQS~\cite{Shammah2018OpenInvariance}, but there the cavity is not included in the reduction, which leads to a larger state space.

\subsection{Polarization orders}
\label{sec:hierarchy}
Within the $c=0$ sector, i.e., fulfilling Eq.~\eqref{eq:u1}, we sort the matrix elements of $\rho$ by their polarization order
\begin{equation}
    b = |n_p - n_p'| = |m - m'| ,
    \label{eq:bandwidth}
\end{equation}
which represents the number of photons involved in the corresponding coherence. The $b=0$ elements thus correspond to the populations, since $n_p=n_p'$ and $m=m'$, whereby one only considers the diagonal of the density matrix. The $b=1$ elements are the optical polarization (first order), which couples to the $b=0$ elements through the Hamiltonian. This can be seen from $H \rho_{n_p j m;\, n_p j m} \propto \rho_{n_p-1 j m+1;\, n_p j m} + \rho_{n_p+1 j m-1;\, n_p j m}$; thus applying the Hamiltonian to a population creates elements with $b=1$. Similarly, the $b=2$ elements are two-photon coherences (second order), which couple to $b=1$ through the Hamiltonian, and so on. We formalize this by defining the orthogonal projectors $\mathcal{P}_b$ that keep the elements of polarization order $b$ and discard the rest, i.e., $\mathcal{P}_b \rho_{n_p j m;\, n_p' j m'} = \delta_{b,\abs{n_p-n_p'}} \rho_{n_p j m;\, n_p' j m'}$. We have $\sum_b \mathcal{P}_b = 1$ on the $c=0$ sector, and we define the first three polarization orders as
\begin{equation}
    P = \mathcal{P}_0\,\rho, \qquad
    Q = \mathcal{P}_1\,\rho, \qquad
    R = \mathcal{P}_2\,\rho ,
    \label{eq:pqr}
\end{equation}
for the populations $P$, the polarization $Q$, and the two-photon coherences $R$. We note that $P$, $Q$, and $R$ are parts of the density operator $\rho$, and we identify each with the collection of its nonzero matrix elements. The elements of $P$ are the diagonal elements $P(n_p,j,m) = \rho_{n_p j m;\, n_p j m}$, which are real, nonnegative state probabilities. The elements of $Q$ are defined as
$Q(n_p,j,m) = \rho_{n_p j m;\, n_p+1, j, m-1}$ and represent coherences between states differing by the exchange of one excitation between the emitter ensemble and the cavity mode. Similarly, the elements of $R$ are defined as
$R(n_p,j,m) = \rho_{n_p j m;\, n_p+2, j, m-2}$ and represent coherences between states differing by two such exchanges. The elements of $Q$ and $R$ are generally complex coherence amplitudes. Their Hermitian-conjugate matrix elements are not independent, since $\rho_{n_p+1, j, m-1;\, n_p j m} = (\rho_{n_p j m;\, n_p+1, j, m-1})^* = Q(n_p,j,m)^*$ and $\rho_{n_p+2, j, m-2;\, n_p j m} = (\rho_{n_p j m;\, n_p+2, j, m-2})^* = R(n_p,j,m)^*$.
When we write equations of motion such as Eq.~\eqref{eq:chain} below, the blocks $\Lb{b}{b'}$ are understood as the matrices that act on these collections of elements, arranged as column vectors.

To derive the dynamics of these polarization orders, we first note two selection rules for the parts of the Liouvillian. First, both terms of the Hamiltonian in Eq.~\eqref{eq:hamiltonian} change the photon number by exactly one, on either the ket or the bra side, so $\mathcal{L}_H$ shifts the polarization order by exactly one:
\begin{equation}
    \mathcal{P}_b\, \mathcal{L}_H\, \mathcal{P}_{b'} = 0 \quad\text{unless}\quad |b - b'| = 1 .
    \label{eq:selection_H}
\end{equation}
In particular, $\mathcal{P}_b \mathcal{L}_H \mathcal{P}_b = 0$: the Hamiltonian has no action within a polarization order. Second, every dissipator shifts the ket and bra indices together, so $\mathcal{L}_D$ is block diagonal:
\begin{equation}
    \mathcal{P}_b\, \mathcal{L}_D\, \mathcal{P}_{b'} = 0 \quad\text{unless}\quad b = b' .
    \label{eq:selection_D}
\end{equation}
Inserting $\sum_b \mathcal{P}_b = 1$ into the master equation, the two selection rules leave only nearest-neighbor couplings between the polarization orders,
\begin{equation}
    \frac{d}{dt}\, \mathcal{P}_b\rho
    = \Lb{b}{b}\, \mathcal{P}_b \rho
    + \Lb{b}{b-1}\, \mathcal{P}_{b-1}\rho
    + \Lb{b}{b+1}\, \mathcal{P}_{b+1}\rho ,
    \label{eq:chain}
\end{equation}
where we defined $\Lb{b}{b'} = \mathcal{P}_b \mathcal{L} \mathcal{P}_{b'}$. Furthermore, the selection rules give $\mathcal{P}_b \mathcal{L} \mathcal{P}_b = \mathcal{P}_b \mathcal{L}_D \mathcal{P}_b$ and $\mathcal{P}_b \mathcal{L} \mathcal{P}_{b\pm1} = \mathcal{P}_b \mathcal{L}_H \mathcal{P}_{b\pm1}$. Thus, the dynamics is a hopping chain in the polarization order, in which the diagonal blocks $\Lb{b}{b}$ are purely dissipative and the off-diagonal blocks are purely Hamiltonian. Ordering the basis by polarization order, we can write the Liouvillian (up to $b=2$) as
\begin{equation}
  \mathcal{L} =
  \begin{pmatrix}
    \Lb{0}{0} & \Lb{0}{1} & 0 \\
    \Lb{1}{0} & \Lb{1}{1} & \Lb{1}{2} \\
    0      & \Lb{2}{1} & \Lb{2}{2}
  \end{pmatrix} ,
  \label{eq:blocks}
\end{equation}
where the vanishing corners are exact: populations couple to two-photon coherences only \emph{through} the polarization.

\begin{figure}[!ht]
    \centering
    \includegraphics[width=\linewidth]{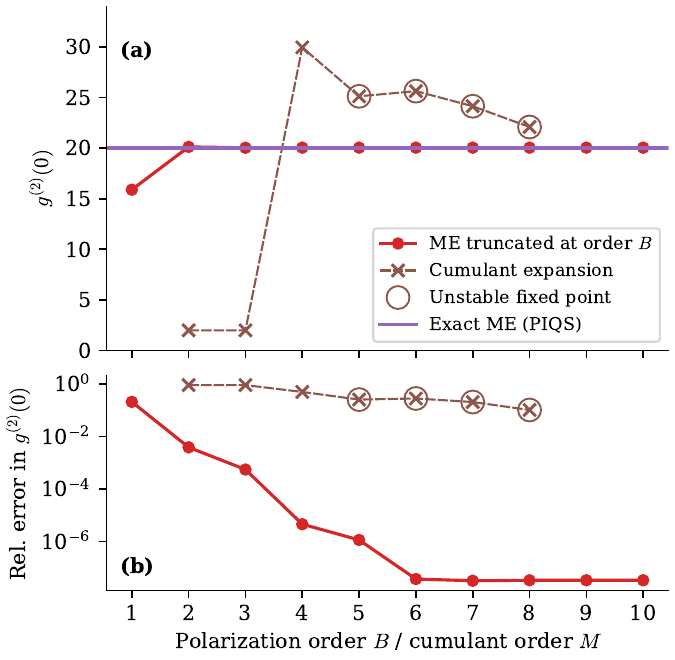}
    \caption{Convergence of the polarization-order hierarchy of Sec.~\ref{sec:hierarchy}, compared with the cumulant expansion, for the strongly collective test case with $n_0 = 10$ emitters, $g = 0.1\,\mathrm{ps}^{-1}$, $\gamma_c = 0.04\,\mathrm{ps}^{-1}$, $\gamma_P = 10^{-4}\,\mathrm{ps}^{-1}$, and $\gamma_A = \gamma_D = 0$. (a) $g^{(2)}(0)$ of the master equation truncated at polarization order $B$ (red) and of the cumulant expansion truncated at order $M$ (brown crosses), compared with the exact ME solved with PIQS (purple). For $M \ge 5$, the steady state of the truncated equations is linearly unstable (circled). (b) Relative error versus $B$ and $M$. For $B=10$, the solution of the symmetry-reduced ME agrees with the PIQS solution to within numerical precision.}
    \label{fig:bandwidth}
\end{figure}

With this formulation, we can introduce the truncation order $B$ as a numerical parameter by setting $\mathcal{P}_b \rho = 0$ for all $b > B$, which determines up to which polarization order coherences are included. We note that since $|n_p - n_p'| = |m - m'| \le 2j \le n_0$, no element is discarded for $B \ge n_0$, and one has a numerically exact description. The truncation order can thus be used as a convergence parameter, which is increased until the changes in the results are insignificant. We use this as the numerically exact reference for all benchmarks.

\subsection{Numerical convergence in the polarization order \label{sec:num_coh_convergence}}

We now consider the convergence in the truncation order $B$ for a strongly collective case of $n_0 = 10$ emitters, with parameters representative of quantum emitters in photonic-crystal cavities, as in Refs.~\cite{Bundgaard-Nielsen2023StochasticNanolaser,Bundgaard-Nielsen2025SimpleNanolasers}: $g = 0.1\,\mathrm{ps}^{-1}$ \cite{Nomura2010LaserSystem} and $\gamma_c = 0.04\,\mathrm{ps}^{-1}$ \cite{Dimopoulos2022Electrically-DrivenThreshold}. To isolate the collective physics, we set $\gamma_A = \gamma_D = 0$, making this a highly coherent test case, and we consider a small pump rate of $\gamma_P = 10^{-4}\,\mathrm{ps}^{-1}$.

Figure~\ref{fig:bandwidth}(a) shows the second-order autocorrelation function
\begin{equation}
g^{(2)}(0)
=
\frac{\expval{a^\dagger a^\dagger a a}}
{\expval{a^\dagger a}^{2}}
\end{equation}
for different truncation orders $B$, see Appendix \ref{app:reference} for numerical details. For comparison, we also show the result obtained with the exact PIQS solver~\cite{Shammah2018OpenInvariance,Johansson2012QuTiP}, which builds the full Liouvillian expanded in Dicke states but does not truncate the polarization order. Keeping only the first polarization order, $B=1$, already captures a clear superradiant signature, with $g^{(2)}(0)\approx16$. Including the second polarization order increases this value to $g^{(2)}(0)\approx20.1$, in close agreement with the exact value of $20.04$. Increasing the truncation order from $B=1$ to $B=2$ thus removes almost all of the deviation from the exact result.

At $B=1$, the only coherence left is the polarization $Q$, and, as we show in Sec.~\ref{sec:lmc}, adiabatically eliminating it leads to a Markov chain, which inherits nearly the same accuracy and thus predicts superradiance with a jump process with modified rates. Much of the superradiant behavior can therefore be understood from modified transition rates at first order. Expanding to second order introduces the two-photon coherences $R$ and the corresponding multiphoton processes, which account for nearly all of the remaining difference from the exact result.

Without truncation, $B \ge n_0$, the symmetry-reduced implementation agrees with the PIQS master-equation solution. Unlike PIQS, the present implementation also uses the charge-conserving restriction ($c=0$) and allows the truncation order to be varied as a numerical parameter. In Sec.~\ref{sec:results_cost}, we compare the numerical cost of PIQS with that of the symmetry-reduced approach. In the next section, we use the polarization-order hierarchy to derive the Dicke LMC and the population--polarization model.

For comparison, Fig.~\ref{fig:bandwidth} also shows the cumulant expansion (Sec.~\ref{sec:results}) truncated at order $M$. For $M \le 3$, the closure fixes $g^{(2)}(0) = 2$, and at $M = 4$, the lowest order that can describe bunching, it overestimates the exact value by about $50\%$. Higher orders reduce the error only slowly, to about $10\%$ at $M = 8$, and for $M \ge 5$ the steady state of the truncated equations is linearly unstable and thus cannot be reached via dynamical evolution (instead found by Newton's method, see appendix \ref{sec:methods_cumulant}). Pushing the order further is also impractical: while the number of equations grows only from $18$ at $M = 4$ to $103$ at $M = 8$, the symbolic derivation of these equations takes $\sim 30$~min at $M = 7$ and $15$~h at $M = 8$ (Appendix~\ref{sec:methods_cumulant}). The polarization-order hierarchy, in contrast, truncates a linear equation, so a steady state always exists, the error drops by about an order of magnitude per step, and $B \ge n_0$ is exact.

\section{Laser models by adiabatic elimination of higher polarization orders}
\label{sec:elimination}
In the previous section, we derived the equations of motion for the polarization orders and introduced the truncation order $B$. In this section, we show that by adiabatically eliminating the polarization in the $B = 1$ truncation, one obtains a formulation in which only probabilities, i.e., the diagonal elements of $\rho$, are present. This formulation is of the same type as the LMC and the stochastic methods used to describe laser dynamics in Refs.~\cite{Roy-Choudhury2009QuantumLasers,Roy-Choudhury2010QuantumDiodes,Lebreton2013StochasticallyNanolasers,Puccioni2015StochasticLasing,Mork2018RateEmitters,Takemura2019LasingLasers,Andre2020,Lippi2021AmplifiedNanolasers,Bundgaard-Nielsen2023StochasticNanolaser,Puccioni2024StatisticsNanolasers,Bundgaard-Nielsen2025SimpleNanolasers,Bundgaard-Nielsen2026ModelingChains}, where the emitter and photon populations are integer valued and governed by jump processes. In the probabilistic description derived here, the Dicke quantum numbers $j$ and $m$ instead undergo quantized jumps together with the photon number $n_p$. Explicitly tracking $j$ allows the model to distinguish collective-spin states with different radiative properties and thereby capture superradiance within a simple stochastic description, which we refer to as the Dicke LMC; see also Fig.~\ref{fig:schematic}. The conventional LMC, which only keeps track of the photon number $n_p$ and the number of excited emitters $n_e$, is recovered when the $j$-resolution is dropped, as shown in Appendix~\ref{app:lmc_limit}.

Starting instead from the $B=2$ truncation and adiabatically eliminating the two-photon coherences $R$ yields the population--polarization model discussed in Sec.~\ref{sec:secondorder}. This model is considerably more accurate, but it can no longer be interpreted as a classical jump process. We show that interference between successive emissions in the two-photon coherences reduces the rates of regular emission and absorption, and that these rates become negative in superradiant bursts, so that the population equation is not a Markov chain over $(n_p, j, m)$.

\subsection{First order: the Dicke Laser Markov Chain}
\label{sec:lmc}
We first truncate the chain in Eq.~\eqref{eq:chain} at $B=1$ and thus retain only the populations $P$ and the polarization $Q$. After the polarization $Q$ is adiabatically eliminated, this gives a purely probabilistic description. The derivation is similar to that of Ref.~\cite{Bundgaard-Nielsen2026ModelingChains}, except that we here expand in the Dicke states, which allows us to keep emitter--emitter correlations. The truncated equations for $P$ and $Q$ read
\begin{align}
    \dot{P} &= \Lb{0}{0}\, P + \Lb{0}{1}\, Q ,
    \label{eq:block_p}\\
    \dot{Q} &= \Lb{1}{0}\,P + \Lb{1}{1}\,Q .
    \label{eq:block_q}
\end{align}
The populations drive the polarization through the Hamiltonian block $\Lb{1}{0}$ and receive the radiative feedback through $\Lb{0}{1}$: radiative transitions are second-order processes in $g$. Formally integrating Eq.~\eqref{eq:block_q} from a diagonal initial state and making a Markov approximation, $P(s)\to P(t)$ inside the memory integral (Appendix~\ref{app:derivation}), the polarization follows the populations adiabatically,
\begin{equation}
    Q(t) \simeq -\Lb{1}{1}^{-1}\, \Lb{1}{0}\, P(t) .
    \label{eq:adiabatic}
\end{equation}
The dissipative block $\Lb{1}{1}$ acts on the polarization elements generated by the Hamiltonian, which all lie in the collective polarization sector spanned by $a^\dagger J_-\rho$ and its conjugates. Instead of calculating $\Lb{1}{1}^{-1}$ explicitly, we use the fact that the collective polarization is an \emph{exact} left eigenoperator of the dissipative Liouvillian,
\begin{equation}
    \mathcal{L}_D^\dagger\left(a^\dagger J_-\right) = -\frac{\gamma_\perp}{2}\, a^\dagger J_-, \qquad
    \gamma_\perp = \gamma_P + \gamma_A + \gamma_D + \gamma_c ,
    \label{eq:eigenoperator}
\end{equation}
as verified term by term in Appendix~\ref{app:derivation}. Thus, we can approximate $\Lb{1}{1}$ by the scalar replacement
\begin{equation}
    \Lb{1}{1} \;\to\; -\frac{\gamma_\perp}{2}\,\mathcal{P}_1 ,
    \label{eq:scalar_closure}
\end{equation}
which discards the internal redistribution of the individual coherence elements while retaining the correct damping of every polarization expectation value. This is the Dicke-basis analog of both the adiabatic elimination of the polarization in laser rate equations \cite{Strauf2011SingleNanolaser,Moelbjerg2013DynamicalEmitters} and the closure step of a cumulant expansion.

The Markov approximation in Eq.~\eqref{eq:adiabatic} and the closure in Eq.~\eqref{eq:scalar_closure} both assume that the polarization is the fastest variable of the system, as for the adiabatic elimination of the polarization in conventional laser rate equations \cite{Strauf2011SingleNanolaser,Moelbjerg2013DynamicalEmitters}. First, the Markov approximation requires that the emitters and the cavity do not exchange excitation coherently before the polarization decays, i.e.,
\begin{equation}
    G_-(n_p,j,m) = g\sqrt{n_p+1}\,\sqrt{C_-(j,m)} \ll \frac{\gamma_\perp}{2}
    \label{eq:validity}
\end{equation}
for the states carrying appreciable probability. Here, $G_-(n_p,j,m) = \bra{n_p{+}1,j,m{-}1} H \ket{n_p,j,m}$ is the collective coupling. Note that it is the Dicke-enhanced coupling, not the bare coupling $g$, that enters, so the condition is most stringent in the superradiant manifolds and at large photon numbers. However, Eq.~\eqref{eq:validity} only concerns the dynamics. In the stationary state, $\dot Q = 0$, and Eq.~\eqref{eq:adiabatic} is exact, so stationary quantities such as $\expval{n_p}$ and $g^{(2)}(0)$ do not require Eq.~\eqref{eq:validity}. Second, cavity loss moves the individual polarization elements to lower photon numbers [Eq.~\eqref{eq:app_qn}], which the closure neglects. This is justified when the polarization decays faster than the cavity field,
\begin{equation}
    \gamma_c \ll \gamma_\perp ,
    \label{eq:validity_closure}
\end{equation}
and this condition also applies in the stationary state.

The test case of Sec.~\ref{sec:num_coh_convergence} violates both conditions below threshold. Since $\gamma_A = \gamma_D = 0$, the polarization decays at the rate $\gamma_\perp = \gamma_P + \gamma_c \approx \gamma_c$, and even the bare coupling $g$ exceeds $\gamma_\perp/2$ by a factor of five. Nevertheless, we show in Sec.~\ref{sec:results} that the Dicke LMC reproduces the mean photon number accurately and captures most of the strong bunching, $g^{(2)}(0) \approx 20$ to within 20\%. We attribute this to the collective factors $C_-(j,m)$, which the Dicke LMC keeps exactly, so that it distinguishes the bright states from the dark states in which the excitation is stored below threshold. Above threshold, the pump increases $\gamma_\perp$, so that Eq.~\eqref{eq:validity_closure} is fulfilled, whereas Eq.~\eqref{eq:validity} is violated even more strongly because of the large photon number. In this regime, the Dicke LMC agrees with the master equation to within about $1\%$ in both $\expval{n_p}$ and $g^{(2)}(0)$, thus showing the Markov approximation is not necessary in the steady state. In contrast, when considering time-dependent quantities such as $g^{(2)}(\tau)$, the Markov approximation in Eq.~\eqref{eq:adiabatic} is important, and here we observe larger deviations, since the Dicke LMC cannot resolve the collective vacuum Rabi oscillations (Sec.~\ref{sec:results_g2tau}).

Inserting Eqs.~\eqref{eq:adiabatic} and \eqref{eq:scalar_closure} into Eq.~\eqref{eq:block_p} gives the closed population equation
\begin{equation}
    \dot{P} = \Lb{0}{0}\,P + \frac{2}{\gamma_\perp}\, \Lb{0}{1}\Lb{1}{0}\, P
    = \Lb{0}{0}\,P - \frac{2}{\gamma_\perp}\, \mathcal{P}_0\comm{H}{\comm{H}{P}} ,
    \label{eq:diagonal_eq}
\end{equation}
where the second form uses that $\mathcal{L}_H$ has no action within a polarization order. Working out the double commutator (Appendix~\ref{app:derivation}) shows that it acts as two effective Lindblad dissipators,
\begin{equation}
    -\frac{2}{\gamma_\perp}\mathcal{P}_0\comm{H}{\comm{H}{\,\cdot\,}} = \gamma_r\, \mathcal{D}_{a^\dagger J_-}[\cdot] + \gamma_r\, \mathcal{D}_{a J_+}[\cdot] ,
    \label{eq:effective_lindblad}
\end{equation}
where we defined $\gamma_r = 4g^2 / \gamma_\perp$. The two dissipators describe the exchange of quanta between the emitters and the cavity as a pair of classical jump processes. We can thus write the evolution of the probability $P$ as:
\begin{equation}
\begin{aligned}
    \dot{P} &= \gamma_r\, \mathcal{D}_{a^\dagger J_-}[P] + \gamma_r\, \mathcal{D}_{a J_+}[P] + \gamma_c \mathcal{D}_a[P] + \sum_i \gamma_A \mathcal{D}_{\sigma_i}[P] \\
    &+ \sum_i \gamma_P \mathcal{D}_{\sigma_i^\dagger}[P] + \sum_i \gamma_D \mathcal{D}_{\sigma_i^\dagger \sigma_i}[P] .
    \label{eq:diagonal_eq_final}
\end{aligned}
\end{equation}
We note that Eq.~\eqref{eq:diagonal_eq_final} has exactly the form of a classical master equation of birth--death type \cite{Gillespie1976AReactions,Gillespie1992AEquation, Gillespie2007StochasticKinetics},
\begin{equation}
    \frac{\partial P(\mathbf{x},t)}{\partial t} = \sum_{\ell}\left[a_\ell(\mathbf{x}-\mathbf{v}_\ell) P(\mathbf{x}-\mathbf{v}_\ell,t) - a_\ell(\mathbf{x})P(\mathbf{x},t)\right],
    \label{eq:cme}
\end{equation}
for the probabilities $P(\mathbf{x},t) = \bra{\mathbf{x}}P(t)\ket{\mathbf{x}}$ over the classical states $\mathbf{x} = (n_p, j, m)$. The resulting transition rates $a_\ell(\mathbf{x}) \ge 0$ and integer state changes $\mathbf{v}_\ell$ of the events $\ell$ are listed in Table~\ref{tab:events_dicke}. To see that Eq.~\eqref{eq:diagonal_eq_final} takes this form, note that every dissipator in it maps basis states onto basis states: for each event $\ell$ with jump operator $A_\ell$, one has $A_\ell \ket{\mathbf{x}} = \sqrt{a_\ell(\mathbf{x})}\, \ket{\mathbf{x} + \mathbf{v}_\ell}$ up to a phase. For the emitter dissipators (pump, decay, and dephasing), the jump operators branch into different values of the quantum number $j$, so that several events are associated with each dissipator.

Projecting onto the diagonal, the first part of the dissipator gives a gain-like term, $\bra{\mathbf{x}} A_\ell P A_\ell^\dagger \ket{\mathbf{x}} = a_\ell(\mathbf{x} - \mathbf{v}_\ell) P(\mathbf{x} - \mathbf{v}_\ell, t)$, and the anticommutator gives a loss-like term, $-\tfrac{1}{2}\bra{\mathbf{x}} \{A_\ell^\dagger A_\ell, P\} \ket{\mathbf{x}} = -a_\ell(\mathbf{x}) P(\mathbf{x},t)$. In Appendix~\ref{app:rates}, we explicitly calculate the branches, transition vectors, and rates. We refer to this purely classical rate equation for the probabilities as the Dicke Laser Markov Chain (Dicke LMC). In the following, we comment on its events and their physical interpretation.

\begin{table}[!ht]
\centering
\caption{Transition rates of the Dicke Laser Markov Chain. The state is $\mathbf{x} = (n_p, j, m)$, and $\mathbf{v}_\ell$ denotes the change of $(n_p,j,m)$ at each event. The branch prefactors $\eta_{0,\pm}(j)$ are given in Eq.~\eqref{eq:eta} and the squared matrix elements in Appendix~\ref{app:rates}. Transitions to nonexistent Dicke states carry zero rate.
\label{tab:events_dicke}}
\begin{ruledtabular}
\begin{tabular}{lcc}
\textbf{Event} & \textbf{Rate $a_\ell$} & $\mathbf{v}_\ell \ (n_p,j,m)$ \\
\colrule
Collective emission & $\gamma_r (n_p{+}1)\, C_-(j,m)$ & $(+1,0,-1)$ \\
Collective absorption & $\gamma_r n_p\, C_+(j,m)$ & $(-1,0,+1)$ \\
Cavity decay & $\gamma_c n_p$ & $(-1,0,0)$ \\
Decay, $j\to j$ & $\gamma_A\, \eta_0(j) |A^{j,m}_-|^2$ & $(0,0,-1)$ \\
Decay, $j\to j{-}1$ & $\gamma_A\, \eta_-(j) |B^{j,m}_-|^2$ & $(0,-1,-1)$ \\
Decay, $j\to j{+}1$ & $\gamma_A\, \eta_+(j) |D^{j,m}_-|^2$ & $(0,+1,-1)$ \\
Pump, $j\to j$ & $\gamma_P\, \eta_0(j) |A^{j,m}_+|^2$ & $(0,0,+1)$ \\
Pump, $j\to j{-}1$ & $\gamma_P\, \eta_-(j) |B^{j,m}_+|^2$ & $(0,-1,+1)$ \\
Pump, $j\to j{+}1$ & $\gamma_P\, \eta_+(j) |D^{j,m}_+|^2$ & $(0,+1,+1)$ \\
Dephasing, $j\to j{-}1$ & $\gamma_D\, \eta_-(j) |B^{j,m}_z|^2$ & $(0,-1,0)$ \\
Dephasing, $j\to j{+}1$ & $\gamma_D\, \eta_+(j) |D^{j,m}_z|^2$ & $(0,+1,0)$ \\
\end{tabular}
\end{ruledtabular}
\end{table}

The rates coming from the radiative dissipators $\gamma_r\, \mathcal{D}_{a^\dagger J_-}[P]$ and $\gamma_r\, \mathcal{D}_{a J_+}[P]$ inherit the Dicke factors of Eq.~\eqref{eq:cpm}: collective emission occurs at the rate $\gamma_r (n_p + 1) C_-(j,m)$, containing both stimulated ($\propto n_p$) and spontaneous ($\propto 1$) contributions, and collective absorption at the rate $\gamma_r n_p C_+(j,m)$. These events move the system along the same $j$ ladder while changing the photon number $n_p$ and the inversion $m$ by $\pm 1$, as illustrated in Fig.~\ref{fig:schematic}(e) and listed in Table~\ref{tab:events_dicke}.

The local dissipators, in contrast, couple different $j$ ladders with each other. A local jump operator such as $\sigma_i$ connects a Dicke state to the three neighboring ladders $j' \in \{j, j-1, j+1\}$, with branching weights known in closed form \cite{Chase2008CollectiveParticles,Baragiola2010CollectiveSystems,Shammah2018OpenInvariance}. Using the branch prefactors
\begin{equation}
\begin{aligned}
    \eta_0(j) &= \frac{1}{2}\,\frac{S+1}{j(j+1)}, \qquad
    \eta_-(j) = \frac{1}{2}\,\frac{S+j+1}{j(2j+1)}, \\
    \eta_+(j) &= \frac{1}{2}\,\frac{S-j}{(j+1)(2j+1)},
    \label{eq:eta}
\end{aligned}
\end{equation}
with $S = n_0/2$, and the squared Dicke matrix elements $|A^{j,m}_q|^2$, $|B^{j,m}_q|^2$, $|D^{j,m}_q|^2$ given in Appendix~\ref{app:rates}, the pump, decay, and dephasing events acquire the rates listed in Table~\ref{tab:events_dicke}. Their sums over the three branches recover the total single-emitter rates, e.g., $a_{A,0} + a_{A,-} + a_{A,+} = \gamma_A n_e$, so that no dissipation is lost by the change of basis. The key new physics is that pure dephasing couples different $j$ ladders with each other, and is a new type of event that was not present in the conventional LMC.

Finally, we note that if the $j$-resolution is discarded and one only tracks $n_p$ and the number of excited emitters $n_e = m + n_0/2$, one recovers the conventional LMC, sketched in Fig.~\ref{fig:schematic}(d). The Dicke factors in the radiative rates are replaced by their uncorrelated-emitter counterparts, $C_-(j,m) \to n_e$ and $C_+(j,m) \to n_0 - n_e$, the local pump and decay branches collapse onto single events with total rates $\gamma_P (n_0 - n_e)$ and $\gamma_A n_e$, and pure dephasing drops out entirely, since it only moves the ensemble between $j$-ladders. The event-by-event correspondence is derived in Appendix~\ref{app:lmc_limit} and summarized in Table~\ref{tab:lmc_limit}.

What is lost in this limit can be seen at the operator
level. Writing $J_+J_- = \sum_i\sigma_i^\dagger\sigma_i + \sum_{i\neq k}\sigma_i^\dagger\sigma_k
= n_e + \sum_{i\neq k}\sigma_i^\dagger\sigma_k$ and using
$C_-(j,m)=\bra{j,m}J_+J_-\ket{j,m}$, we have
\begin{equation}
    C_-(j,m) = n_e + \sum_{i\neq k}\bra{j,m}\sigma_i^\dagger\sigma_k\ket{j,m} ,
    \label{eq:cminus_pairwise}
\end{equation}
and likewise $C_+(j,m) = (n_0-n_e) + \sum_{i\neq k}\bra{j,m}\sigma_i\sigma_k^\dagger\ket{j,m}$.
The first term of Eq.~\eqref{eq:cminus_pairwise} depends on $m$ alone. The entire $j$-dependence, and hence the collective enhancement or suppression of the radiative rate, is thus carried by the pairwise interemitter coherence $\langle\sigma_i^\dagger\sigma_k\rangle$, which is neglected when the $j$-dependence is discarded.

\subsection{Second order: the population--polarization model}
\label{sec:secondorder}
We now repeat the elimination one polarization order higher. Truncating the chain in Eq.~\eqref{eq:chain} at $B=2$ adds the two-photon coherences $R$,
\begin{align}
    \dot{P} &= \Lb{0}{0}\, P + \Lb{0}{1}\, Q ,\nonumber\\
    \dot{Q} &= \Lb{1}{0}\, P + \Lb{1}{1}\, Q + \Lb{1}{2}\, R ,
    \label{eq:b2_system}\\
    \dot{R} &= \Lb{2}{1}\, Q + \Lb{2}{2}\, R \nonumber,
\end{align}
where the absence of a direct $P\leftrightarrow R$ coupling is exact, since the Hamiltonian only couples polarization orders that differ by one. Adiabatically eliminating the highest order, $R \simeq -\Lb{2}{2}^{-1}\Lb{2}{1}\,Q$, again requires an estimate of $\Lb{2}{2}^{-1}$. However, the closure argument of Sec.~\ref{sec:lmc} carries over: the collective two-photon coherence operator is again an exact left eigenoperator of the dissipative Liouvillian,
\begin{equation}
    \mathcal{L}_D^\dagger\!\left[(a^\dagger)^2 J_-^2\right] = -\gamma_\perp\, (a^\dagger)^2 J_-^2 ,
    \label{eq:eigenoperator2}
\end{equation}
now with twice the polarization eigenvalue $\gamma_\perp/2$ (Appendix~\ref{app:derivation}). The corresponding scalar closure, $\Lb{2}{2} \to -\gamma_\perp\,\mathcal{P}_2$, allows us to eliminate the second polarization order and insert into the equation for the first order, giving the closed system
\begin{equation}
    \;
    \dot{P} = \Lb{0}{0}\, P + \Lb{0}{1}\, Q , \qquad
    \dot{Q} = \Lb{1}{0}\, P + F\, Q ,
    \;
    \label{eq:resum_model}
\end{equation}
with
\begin{equation}
    F = \Lb{1}{1} + \frac{1}{\gamma_\perp}\, \Lb{1}{2}\Lb{2}{1} .
    \label{eq:folded}
\end{equation}
Here, $\Lb{1}{1}$ is no longer approximated but is included exactly. We refer to Eq.~\eqref{eq:resum_model} as the \emph{population--polarization model}: it explicitly includes the polarization $Q$ in its dynamics, while the two-photon coherences are folded in through the exact eigenvalue relation in Eq.~\eqref{eq:eigenoperator2}. We note that, unlike its first-order parent, the model is no longer a classical jump process, since the polarization variables $Q$ are complex amplitudes and not probabilities. It does, however, still drastically reduce the state space compared with the full master equation, as the state space only grows by the $b=1$ elements (a factor of $\sim 3$).

As in the first-order expansion, the elimination can formally be carried through to a population-only model. Solving Eq.~\eqref{eq:resum_model} for $Q$ at stationarity gives $Q = -F^{-1} \Lb{1}{0} P$ and thus $\dot P = \mathcal{L}_{\mathrm{eff}} P$, with $\mathcal{L}_{\mathrm{eff}} = \Lb{0}{0} - \Lb{0}{1} F^{-1} \Lb{1}{0}$. The resulting equation is, however, not a Markov chain, and the reason is physical.

To see this, we expand $F^{-1}$ in powers of $g$, assuming for simplicity that $\Lb{1}{1} \approx -(\gamma_\perp/2)\,\mathcal{P}_1$:
\begin{equation}
\begin{aligned}
    F^{-1} &\approx \Lb{1}{1}^{-1}
-\frac{1}{\gamma_\perp}
\Lb{1}{1}^{-1}\Lb{1}{2}\Lb{2}{1}
\Lb{1}{1}^{-1}
+\mathcal O\!\left(\frac{g^4}{\gamma_\perp^5}\right) \\
&\approx -\frac{2}{\gamma_\perp}\,\mathcal{P}_1 - \frac{4}{\gamma_\perp^3}\,\Lb{1}{2}\Lb{2}{1} .
\end{aligned}
\end{equation}
Inserting this result into $\mathcal{L}_{\mathrm{eff}}$ yields
\begin{equation}
\begin{aligned}
\mathcal L_{\mathrm{eff}}
\simeq{} &
\Lb{0}{0} + \frac{2}{\gamma_\perp} \Lb{0}{1}\Lb{1}{0}
\\ &+ \frac{4}{\gamma_\perp^3} \Lb{0}{1}\Lb{1}{2}\Lb{2}{1}\Lb{1}{0} + \mathcal O\left(\frac{g^6}{\gamma_\perp^5}\right). \label{eq:leff_expansion}
\end{aligned}
\end{equation}
The first correction, $\frac{2}{\gamma_\perp} \Lb{0}{1}\Lb{1}{0}$, is the radiative term of the Dicke LMC, which is of order $g^2$. It describes the excursion $P\longrightarrow Q\longrightarrow P$ and, after projection onto the populations, can be written as a sum of Lindblad dissipators with positive rates.

The next term is of order $g^4$ and describes an excursion into the second polarization order, i.e., the two-photon coherences,
\begin{equation}
P\longrightarrow Q\longrightarrow R
\longrightarrow Q\longrightarrow P.
\end{equation}
It contains four actions of the Hamiltonian and can be written as
\begin{equation}
\begin{aligned}
\Lb{0}{1}\Lb{1}{2}\Lb{2}{1}\Lb{1}{0}
={}&
\mathcal P_0\mathcal L_H\mathcal P_1
\mathcal L_H\mathcal P_2
\mathcal L_H\mathcal P_1
\mathcal L_H\mathcal P_0 .
\label{eq:fourth_order_excursion}
\end{aligned}
\end{equation}
Using the nearest-neighbor selection rule, this can equivalently be expressed as
\begin{equation}
\begin{aligned}
\Lb{0}{1}\Lb{1}{2}\Lb{2}{1}\Lb{1}{0}
={}&
\mathcal P_0\mathcal L_H^4\mathcal P_0
-
\left(
\mathcal P_0\mathcal L_H^2\mathcal P_0
\right)^2 .
\label{eq:connected_fourth_order}
\end{aligned}
\end{equation}

Evaluating Eq.~\eqref{eq:connected_fourth_order} on a population shows that the excursion through the two-photon coherences has two effects. We consider the states $x_k = (n_p+k,j,m-k)$ along the emission cascade and denote by $\Gamma_k = \gamma_r(n_p+k+1)\,C_-(j,m-k)$ the rate of regular emission from $x_k$ to $x_{k+1}$ in the Dicke LMC. The fourth-order term then adds the rates
\begin{equation}
\begin{aligned}
    x_0 \to x_2: &\quad +\frac{\Gamma_0\Gamma_1}{2\gamma_\perp} ,\\
    x_0 \to x_1: &\quad -\frac{\Gamma_0\left(\Gamma_{-1}+\Gamma_1\right)}{\gamma_\perp} ,
\end{aligned}
\label{eq:fourth_order_rates}
\end{equation}
and the corresponding rates for absorption. The first is a two-photon emission. The second is a negative correction to the regular emission, and it is a coherence effect. We note that each further order of the expansion of $\mathcal{L}_{\mathrm{eff}}$ in Eq.~\eqref{eq:leff_expansion} in powers of $g$ adds higher order corrections.
The two-photon coherence between $x_0$ and $x_2$ is created from the polarizations of both emissions, $x_0 \to x_1$ and $x_1 \to x_2$. The two successive emissions therefore interfere destructively, which reduces the regular emission. Adding this correction to $\Gamma_0$, the rate of regular emission from $x_0$ becomes
\begin{equation}
    \Gamma_0\left(1 - \frac{\Gamma_{-1}+\Gamma_1}{\gamma_\perp}\right) ,
    \label{eq:kab_expansion}
\end{equation}
which is negative once the neighboring emissions are faster than the decay of the polarization, $\Gamma_{-1}+\Gamma_1 \gtrsim \gamma_\perp$. This is the case in a superradiant burst. For example, for the test case in Sec.~\ref{sec:num_coh_convergence} with $n_0 = 10$, the regular emission from $(n_p,j,m) = (0,1,1)$ has $\Gamma_{-1} = 0$ (no previous emission), $\Gamma_0 = 2\gamma_r$ (the emission itself), and $\Gamma_1 = 4\gamma_r$ (the following emission, from $(1,1,0)$ to $(2,1,-1)$). Below threshold, $\gamma_\perp \approx \gamma_c = 0.04\,\mathrm{ps}^{-1}$ and $\gamma_r \approx 1.0\,\mathrm{ps}^{-1}$, so that $\Gamma_1 \approx 100\,\gamma_\perp$,  which is far outside the validity of Eq.~\eqref{eq:kab_expansion} (we need higher order terms). Still, evaluating $\mathcal{L}_{\mathrm{eff}}$ numerically with $\Lb{1}{1}$ replaced by a scalar as above, we find that the higher-order terms correct this value but still lead to a negative rate of $-0.63\,\mathrm{ps}^{-1}$, compared with $\Gamma_0 \approx 2.0\,\mathrm{ps}^{-1}$.

A negative rate is not allowed in a Markov chain, so $\mathcal{L}_{\mathrm{eff}}$ cannot be interpreted as a jump process and cannot be sampled with the GFRM.

For this reason, and since it is also numerically favorable, we never form the dense $\mathcal{L}_{\mathrm{eff}}$. Instead, we solve the sparse system in Eq.~\eqref{eq:resum_model} with the polarization kept explicit. This is faster than forming the dense $\mathcal{L}_{\mathrm{eff}}$, and it has the added benefit that $Q$ need not be adiabatically eliminated, which gives a higher accuracy for time-dependent quantities such as $g^{(2)}(\tau)$. In the next section, we benchmark the Dicke LMC and the population--polarization model against the exact master equation.

\section{Results}
\label{sec:results}
In the previous section, we derived the Dicke LMC and the population--polarization model from the symmetry-reduced master equation. In this section, we compare the reference ME solution (obtained from the symmetry-reduced master equation of Sec.~\ref{sec:model}), the conventional $(n_p,n_e)$ LMC, the Dicke LMC of Sec.~\ref{sec:lmc}, the population--polarization model of Sec.~\ref{sec:secondorder}, and finally the cumulant expansion of QuantumCumulants.jl \cite{Plankensteiner2022quantumcumulantsjl} (see Table~\ref{tab:methods} for an overview). Going from the master equation via the population--polarization model to the Dicke LMC, each step removes coherence between the emitters and the light, while the conventional LMC in addition discards the emitter--emitter correlations. As in Sec.~\ref{sec:num_coh_convergence}, we use $g = 0.1\,\mathrm{ps}^{-1}$ and $\gamma_c = 0.04\,\mathrm{ps}^{-1}$ and, unless stated otherwise, $n_0 = 10$ emitters. To isolate the collective physics, we set $\gamma_A = \gamma_D = 0$, so that the only processes competing with the collective light--matter interaction are the incoherent pump and the cavity loss.

Before presenting the comparisons, we note that the two Markov chains (LMC and Dicke LMC) can be solved in two ways. For stationary observables, one may solve directly for the steady-state probability distribution of the chain. Writing the probabilities as a vector $x$, the dynamics can be expressed as $\dot{x} = M x$, where $M$ is the generator of the Markov chain (see Eq.~\eqref{eq:cme}). The steady state is then obtained by solving the linear system $M x = 0$, together with the normalization condition on $x$. Alternatively, one may generate many individual stochastic trajectories using Gillespie's first reaction method (GFRM) and estimate the observables by averaging over them. The first approach gives the stationary result without statistical sampling error, but becomes increasingly demanding as the number of accessible states grows. The trajectory approach avoids storing and solving for the full stationary distribution, but introduces a statistical error that decreases only as more trajectories are sampled.

The two approaches solve the same model and agree within sampling accuracy. All LMC and Dicke LMC results below are direct stationary solves, except for the largest emitter numbers in Figs.~\ref{fig:g2_vs_emitters} and \ref{fig:benchmark}, where GFRM sampling (labeled ``Dicke LMC (GFRM)'') is the only feasible option and shows the reach of the method. The ME reference and the population--polarization model are always solved directly, although Monte Carlo wave-function sampling \cite{Mlmer1993MonteOptics, Zhang2018MonteLasing} would be the quantum analog of the GFRM.

For the cumulant expansion, QuantumCumulants.jl derives the equations of motion for the expectation values of products of the photon and emitter operators, using indexed emitter operators and the built-in symmetry of identical emitters \cite{Plankensteiner2022quantumcumulantsjl}. The hierarchy is closed at order $M$ by setting all joint cumulants of more than $M$ operators to zero, i.e., by factorizing higher-order expectation values into products of lower-order ones. We use $M = 4$, which is the lowest order at which bunching can appear, throughout (except for Fig.~\ref{fig:bandwidth}, where we consider the convergence). For $M \le 3$ the factorization gives $\expval{a^\dagger a^\dagger a a} = 2\expval{a^\dagger a}^2$ and thus $g^{(2)}(0) = 2$ identically, for any parameters. Higher orders offer little improvement (Fig.~\ref{fig:bandwidth} and Sec.~\ref{sec:num_coh_convergence}): their error decreases only slowly, their steady state below threshold is unstable, so that it is never reached by the dynamics, and their derivation quickly becomes slow. At $M = 4$, the resulting $18$ coupled nonlinear equations are solved for their steady state (see Appendix~\ref{sec:methods_cumulant}).

\subsection{Transition to lasing}

\begin{figure}
    \centering
    \includegraphics[width=\linewidth]{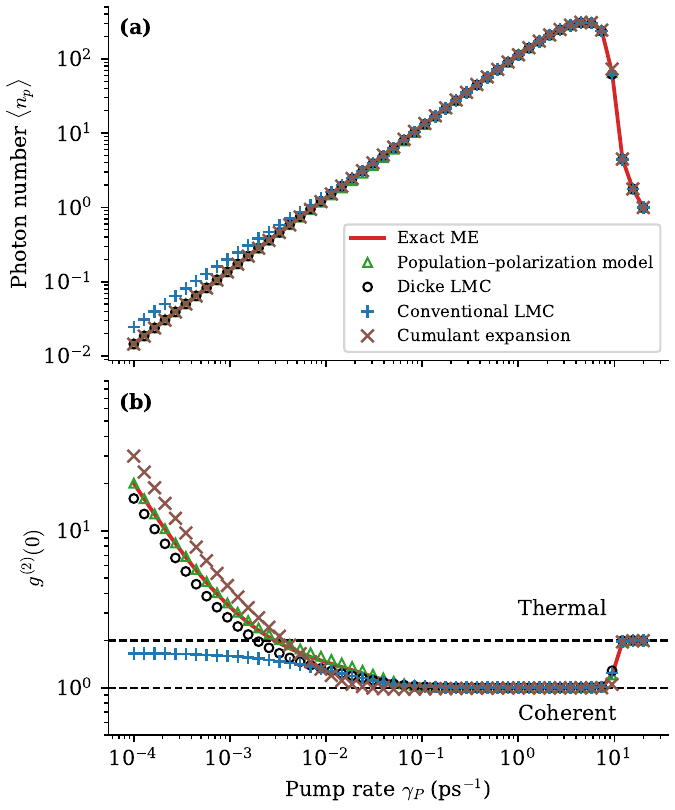}
    \caption{Photon number $n_p$ (a) and second-order correlation function $g^{(2)}(0)$ (b) versus pump rate for $n_0 = 10$ emitters as computed using different models. Parameters are the same as in Fig.~\ref{fig:bandwidth} except for the varied pump rate $\gamma_P$. The Dicke LMC, solved directly as a stationary probability equation (black circles), and the population--polarization model (green triangles) are compared to the numerically exact master equation (red line, Sec.~\ref{sec:num_coh_convergence}), to the conventional LMC over $(n_p, n_e)$ (blue plusses), and to the fourth-order cumulant expansion (brown crosses).}
    \label{fig:pump_sweep}
\end{figure}

Figure~\ref{fig:pump_sweep} shows the mean photon number and $g^{(2)}(0)$ as a function of the pump rate $\gamma_P$ for $n_0 = 10$ emitters and zero dephasing, where collective effects are the strongest. Below threshold, the exact master equation displays collective emission: strongly superthermal photon bunching, reaching $g^{(2)}(0) \approx 20$ at $\gamma_P = 10^{-4}\,\mathrm{ps}^{-1}$, far above the thermal value $g^{(2)}(0)=2$. The physical picture is that of random superradiant bursts \cite{Jahnke2016GiantNanolasers,Leymann2015Sub-Nanolasers}: pumping occasionally prepares the ensemble in a state of large $C_-(j,m)$, which then emits a rapid cascade of photons, while at other times the ensemble is trapped in subradiant states with small $j$ and emits almost nothing. Far above threshold, the decoherence of the pump leads to quenching of emission and a thermal $g^{(2)}(0)=2$, which happens in few-emitter lasers when the pump rate per emitter is large \cite{Bundgaard-Nielsen2023StochasticNanolaser,Bundgaard-Nielsen2025SimpleNanolasers,Bundgaard-Nielsen2026ModelingChains}. 

The Dicke LMC reproduces both the photon number and $g^{(2)}(0)$. The photon number agrees with the master equation to within a few percent over the entire sweep, and the bunching at the lowest pump rate is captured up to a residual underestimate of about $20\%$ ($g^{(2)}(0) = 16.1$ versus $20.0$). This deviation can be explained by the missing two-photon coherences, which, as shown in Sec.~\ref{sec:num_coh_convergence}, substantially improve the accuracy of the truncated master equation. The population--polarization model closes this gap by including the second polarization order. At the lowest pump rate, it yields $g^{(2)}(0) = 20.10$ against the exact $20.04$. Its error grows somewhat toward threshold, where it remains smaller than that of the Dicke LMC, and becomes negligible above threshold. In contrast, the conventional LMC over $(n_p, n_e)$ gives $g^{(2)}(0) = 1.7$ at the lowest pump rate. It does not exceed the thermal value of 2 within the investigated parameter ranges.

Furthermore, the conventional LMC overestimates the photon number because it ignores subradiant trapping, consistent with Ref.~\cite{Leymann2015Sub-Nanolasers}. This can be explained from $ \eta_{\mathrm{sp}}= \expval{C_-(j,m)}/\expval{n_e}$, which by Eq.~\eqref{eq:cminus_pairwise} is below (above) one when the pair coherence $\sum_{i\neq k}\expval{\sigma_i^\dagger\sigma_k}$ is negative (positive). Below threshold, excitations accumulate in subradiant states, and $\eta_{\mathrm{sp}} \ll 1$. With $\gamma_A = 0$ they are not lost, but they occupy emitters that the pump could otherwise excite, which lowers $\expval{n_p}$ (compared with the conventional LMC without these correlations).

Above threshold, the ensemble becomes bright, $\eta_{\mathrm{sp}} > 1$, as in Ref.~\cite{Leymann2015Sub-Nanolasers}, but unlike  Ref.~\cite{Leymann2015Sub-Nanolasers}  we do not see an increase in the photon number (when including coherence effects) in Fig.~\ref{fig:pump_sweep}(a). The increased brightness speeds up emission and reabsorption equally, the two rates differ by $\gamma_r[C_-(j,m) - C_+(j,m)] = 2\gamma_r m$, independent of $j$. As long as the cavity rarely empties, lasing therefore clamps the inversion where this net gain balances the cavity loss, $\expval{m} = \gamma_c/(2\gamma_r)$. Since the photon number is set by how many emitters the pump can still excite, it is the same as for independent emitters. Below threshold, by contrast, the cavity is mostly empty, emitted photons leave before they can be reabsorbed, and brightness controls the output directly. In Ref.~\cite{Leymann2015Sub-Nanolasers}, the coupling is weaker than the cavity loss, $g < \gamma_c$, and the cavity holds few photons, so this compensation is incomplete, and they observe a photon enhancement above threshold.

Superradiance above threshold is also the operating principle of the ultranarrow-linewidth lasers based on cold atoms \cite{Meiser2009ProspectsLaser,Bohnet2012ALaser,Debnath2018LasingRegime}, where it narrows the line below the Schawlow--Townes linewidth $\Delta\omega_{\mathrm{ST}} = \gamma_r\expval{n_e}/(2\expval{n_p})$. When the emitter linewidth is not large compared with the cavity loss rate, part of the phase coherence of the laser is stored in the collective dipole of the emitters, and the linewidth becomes \cite{Kuppens1994Quantum-limitedLaser}
\begin{equation}
    \Delta\omega \simeq \frac{\Delta\omega_{\mathrm{ST}}}{1 + \gamma_c/\gamma_e} ,
    \label{eq:linewidth}
\end{equation}
where $\gamma_e = \gamma_P + \gamma_A + \gamma_D$ is the decay rate of the emitter dipoles. Note that Ref.~\cite{Kuppens1994Quantum-limitedLaser} writes the reduction factor as $(1+\gamma_c/\gamma_e)^{-2}$, relative to the Schawlow--Townes linewidth defined with the cold-cavity loss rate, $\gamma_c n_{\mathrm{sp}}/(2\expval{n_p})$, where $n_{\mathrm{sp}} = \expval{n_e}/(2\expval{m})$ is the population-inversion factor. Our $\Delta\omega_{\mathrm{ST}}$ is instead defined with the rate $\gamma_r = 4g^2/\gamma_\perp$ of spontaneous emission into the cavity mode, which already contains one of the two factors through $\gamma_\perp = \gamma_e + \gamma_c$. The cold-atom lasers operate in the bad-cavity limit $\gamma_c \gg \gamma_e$, where $\Delta\omega \simeq \Delta\omega_{\mathrm{ST}}\gamma_e/\gamma_c$ is set by the emitters rather than the cavity \cite{Meiser2009ProspectsLaser}. The nanolasers considered here operate in the opposite limit. For $\gamma_A = 0$, the excitation balance gives $\gamma_c\expval{n_p} = \gamma_P\expval{n_g} \le \gamma_P n_0$, with $\expval{n_g} = n_0 - \expval{n_e}$ the number of emitters in the ground state, so a photon number well above $n_0$ requires $\gamma_P \gg \gamma_c$. Since the pump also damps the emitter dipoles, $\gamma_e \ge \gamma_P \gg \gamma_c$, and the reduction factor in Eq.~\eqref{eq:linewidth} is close to one. For the parameters of Fig.~\ref{fig:pump_sweep}, $\gamma_c/\gamma_e = 0.4$ at $\gamma_P = 0.1\,\mathrm{ps}^{-1}$ and $0.02$ at $\gamma_P = 2\,\mathrm{ps}^{-1}$, so the superradiant narrowing amounts to $29\%$ and $2\%$, respectively, and becomes negligible far above threshold.

The fourth-order cumulant expansion behaves differently from all the population-based models. Below threshold, it reproduces the photon number, but it overestimates the bunching, with $g^{(2)}(0) = 29.9$ at the lowest pump rate. Its error thus has the opposite sign to that of the Dicke LMC. Just above threshold, it underestimates $g^{(2)}(0)$, and for $\gamma_P \gtrsim 0.3\,\mathrm{ps}^{-1}$ it agrees with the master equation. The factorization underlying the expansion is accurate for states close to a product state, where the emitters are uncorrelated. Below threshold, the ensemble is far from such a state: the excitation spends most of its time stored in subradiant states, and only occasionally reaches a bright state, from which it is released as a burst of photons. The brightness $\expval{J_+J_-}$ that drives the emission is a small residue of the two nearly canceling terms in Eq.~\eqref{eq:cminus_pairwise}, namely the excited-emitter number $\expval{n_e}$ and the pair coherence $\sum_{i\neq k}\expval{\sigma_i^\dagger\sigma_k}$. For the exact steady state at the lowest pump rate, $\expval{n_e} = 4.20$ and the pair coherence is $-4.17$, leaving only $\expval{J_+J_-} = 0.034$. The cumulant expansion approximates the pair coherence by factorized correlations, and an error of just $1\%$ in it already exceeds $\expval{J_+J_-}$ itself. A small error in the pair coherence thus becomes a large error in the radiative rate, leading to the large deviations in the prediction of $g^{(2)}(0)$.

\subsection{Scaling with emitter number}

\begin{figure}
    \centering
    \includegraphics[width=\linewidth]{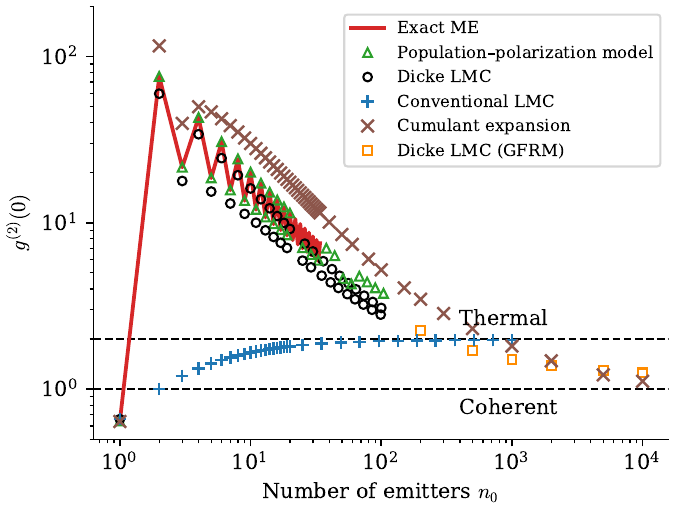}
    \caption{Photon bunching at a fixed low pump rate versus emitter number, with the remaining parameters as in Fig.~\ref{fig:bandwidth} ($\gamma_P = 10^{-4}\,\mathrm{ps}^{-1}$, $\gamma_A = \gamma_D = 0$). Red lines: numerically exact master equation (up to $n_0 = 34$); green triangles: population--polarization model (up to $n_0 = 104$); black circles: stationary solution of the Dicke LMC; orange squares: GFRM sampling of the Dicke LMC at large $n_0$; blue plusses: conventional LMC; brown crosses: fourth-order cumulant expansion. The pronounced even--odd alternation reflects the existence of a fully dark $j = 0$ manifold for even $n_0$ only.}
    \label{fig:g2_vs_emitters}
\end{figure}

Figure~\ref{fig:g2_vs_emitters} shows the bunching at the low pump rate $\gamma_P=10^{-4}\,\mathrm{ps}^{-1}$ as a function of the emitter number, again without dephasing and background decay, $\gamma_D=\gamma_A=0$.

First, we see that $g^{(2)}(0)$ alternates strongly between even and odd $n_0$, as also observed in Ref.~\cite{Auffeves2011FewIndividualization}. For even $n_0$, the system possesses a completely dark $j=0$ manifold in which the pump can trap population, which enhances the sudden bursts of emission. For odd $n_0$, in contrast, $j \ge 1/2$ always holds, and a minimal radiative coupling is thus always present. The effect is largest for few emitters ($g^{(2)}(0) \approx 76$ for $n_0=2$ versus $22$ for $n_0=3$) and washes out slowly with increasing $n_0$, or, as we will see in Sec.~\ref{sec:dephasing} if the pure dephasing is increased. We expect a similar decrease for inhomogeneously broadened emitters, since a frequency difference between two emitters couples the dark singlet to the bright triplet state and thus, like pure dephasing, moves the ensemble between Dicke ladders. Collective emission from inhomogeneously broadened ensembles has been studied in Refs.\cite{Temnov2005SuperradianceCavity,Debnath2019CollectiveLinewidth,Bychek_Hotter_Plankensteiner_Ritsch_2021}. Second, the bunching decays slowly with system size and stays above the thermal value even at $n_0 = 100$, where the direct solve of the Dicke LMC involves $7.8\times10^4$ states and completes in under a minute (see Sec.~\ref{sec:results_cost}). The exact master equation becomes impractical beyond a few tens of emitters ($n_0 = 34$ required $550$~s and was limited by memory). The population--polarization model can be solved up to $n_0 = 104$ and agrees with the master equation over the entire range where the latter is available. GFRM sampling of the Dicke LMC extends the trend to thousands of emitters at an essentially constant cost per step. The Dicke LMC agrees qualitatively with the ME, but it consistently underestimates the bunching, as in Fig.~\ref{fig:pump_sweep}.

The cumulant expansion, in contrast, shows no even--odd alternation. Beyond $n_0 = 4$, its $g^{(2)}(0)$ decreases smoothly with the emitter number. For $n_0 \ge 5$, it therefore overestimates the bunching, in particular for odd $n_0$, while the photon number remains accurate. The reason is that the dark manifold is a joint property of all $n_0$ emitters. The fourth-order expansion keeps correlations among at most four emitters and factorizes all higher ones, and $n_0$ enters its equations only through smooth prefactors such as $n_0-1$. It therefore cannot tell whether $n_0$ is even, i.e., whether a dark manifold exists at all. Consistent with this, the expansion does alternate for $n_0 \le 4$ [$g^{(2)}(0) = 115$, $40$, and $50$ for $n_0 = 2$, $3$, and $4$], where the dark manifold involves at most four emitters.

\subsection{The role of dephasing \label{sec:dephasing}}

\begin{figure}
    \centering
    \includegraphics[width=\linewidth]{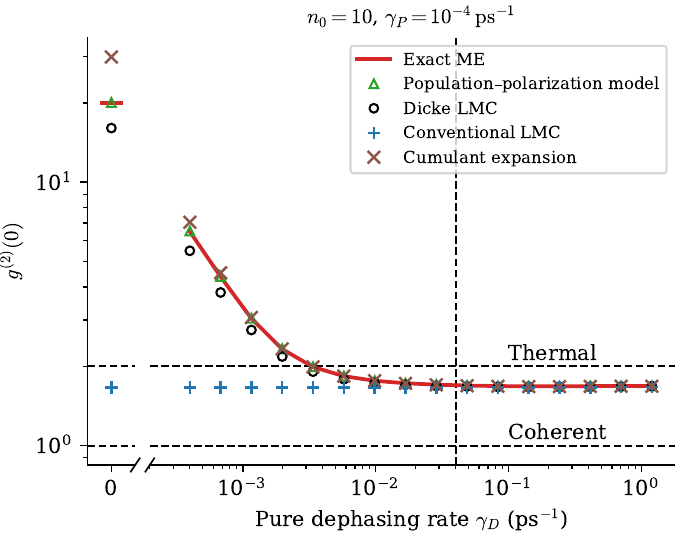}
    \caption{$g^{(2)}(0)$ for $n_0=10$ emitters as a function of the pure dephasing rate $\gamma_D$. Parameters are the same as in Fig.~\ref{fig:bandwidth} ($\gamma_P=10^{-4}\,\mathrm{ps}^{-1}$, $\gamma_A = 0$) except for the varied $\gamma_D$. The exact master equation (red) is compared with the Dicke LMC (black circles), the population--polarization model (green triangles), the conventional LMC (blue plusses), which is essentially unchanged across the sweep, and the fourth-order cumulant expansion (brown crosses). The vertical dashed line marks $\gamma_D = \gamma_c$.
    \label{fig:dephasing_sweep}}
\end{figure}

The results shown so far have considered the fully coherent limit $\gamma_A = \gamma_D = 0$, where the collective effects are largest. Figure~\ref{fig:dephasing_sweep} shows how the picture changes when the rate of pure dephasing is varied. Dephasing scatters the ensemble between Dicke ladder states and destroys the interemitter coherence responsible for the superradiant bursts: $g^{(2)}(0)$ collapses from $20$ toward the uncorrelated value $\approx 1.7$ once $\gamma_D$ exceeds the cavity rate $\gamma_c$ (vertical line). The Dicke LMC tracks this collapse, and its accuracy improves with dephasing. This is in line with Eq.~\eqref{eq:validity_closure}: dephasing increases $\gamma_\perp$, and for $\gamma_D \gtrsim \gamma_c$ the polarization decays before a photon leaves the cavity. The population--polarization model is accurate over the entire sweep. The conventional LMC, in contrast, is almost insensitive to $\gamma_D$, and it only becomes accurate in the strongly dephased regime, where collectivity is absent. The cumulant expansion overestimates the bunching at weak dephasing, but its error drops rapidly, and for $\gamma_D > 2\times10^{-3}\,\mathrm{ps}^{-1}$ it agrees with the master equation more closely than the Dicke LMC. This is expected, as the dephasing drives the ensemble toward an uncorrelated state, where the factorization of the cumulant expansion is justified.

\subsection{Delayed correlations}
\label{sec:results_g2tau}

\begin{figure}
    \centering
    \includegraphics[width=\linewidth]{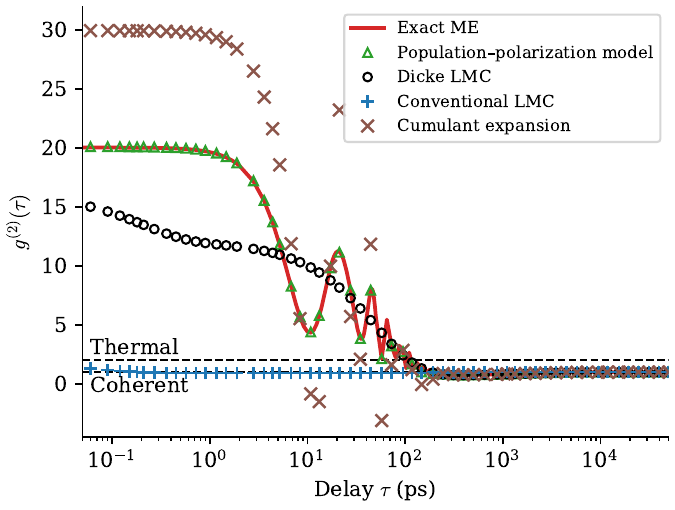}
    \caption{Delayed second-order correlation function $g^{(2)}(\tau)$ in the strongly collective regime. Parameters are the same as in Fig.~\ref{fig:bandwidth} ($n_0 = 10$, $\gamma_P = 10^{-4}\,\mathrm{ps}^{-1}$, $\gamma_A = \gamma_D = 0$). Red line: numerically exact PI master equation, evaluated with the quantum regression theorem in the symmetry-reduced solver; green triangles: population--polarization model; black circles: Dicke LMC via Eq.~\eqref{eq:g2tau}; blue plusses: conventional LMC; brown crosses: fourth-order cumulant expansion, obtained by evolving the conditional state $a\rho a^\dagger/\expval{n_p}$ with the same closed equations. Dashed horizontal lines mark the coherent and thermal values $g^{(2)} = 1$ and $2$.}
    \label{fig:g2_tau}
\end{figure}

Figure~\ref{fig:g2_tau} shows $g^{(2)}(\tau)$ for $\gamma_P=10^{-4}\,\mathrm{ps}^{-1}$, $n_0=10$ emitters, and $\gamma_D=\gamma_A=0$, computed with the master equation and the approximate methods. The numerical details are given in Appendix~\ref{app:numerical_details}.

At short delays, the correlation function $g^{(2)}(\tau)$ predicted by the master equation shows damped collective vacuum Rabi oscillations. Detecting a photon projects the ensemble onto its bright, strongly coupled states, for which the collective coupling $G \sim g\sqrt{C_-}$ exceeds $\gamma_\perp/2$. The emitter ensemble and the cavity thus exchange excitation coherently several times before the coherence decays. This is the regime in which the Markov approximation in Eq.~\eqref{eq:adiabatic} is invalid [Eq.~\eqref{eq:validity}], and the Dicke LMC accordingly averages over the oscillations, relaxing monotonically from its underestimated initial value. The population--polarization model, which keeps the polarization as a dynamical variable, reproduces these oscillations. Beyond $\tau \sim 10^2$~ps, $g^{(2)}(\tau) \approx 1$, and all methods agree.

The conventional LMC, for comparison, shows none of this structure: it drops immediately to $g^{(2)} \approx 1$ and is featureless at all delays. The cumulant expansion starts from its overestimated value $g^{(2)}(0) = 29.9$ and does produce collective Rabi oscillations, but they are strongly exaggerated and swing to unphysical negative values between $10$ and $150$~ps (down to $g^{(2)}(\tau) \approx -3.1$ at $\tau \approx 57$~ps), since nothing in the truncated equations enforces positivity. Beyond $\tau \sim 2\times10^2$~ps, it agrees with the other methods.

\subsection{Computational cost}
\label{sec:results_cost}

\begin{figure*}
    \centering
    \includegraphics[width=\linewidth]{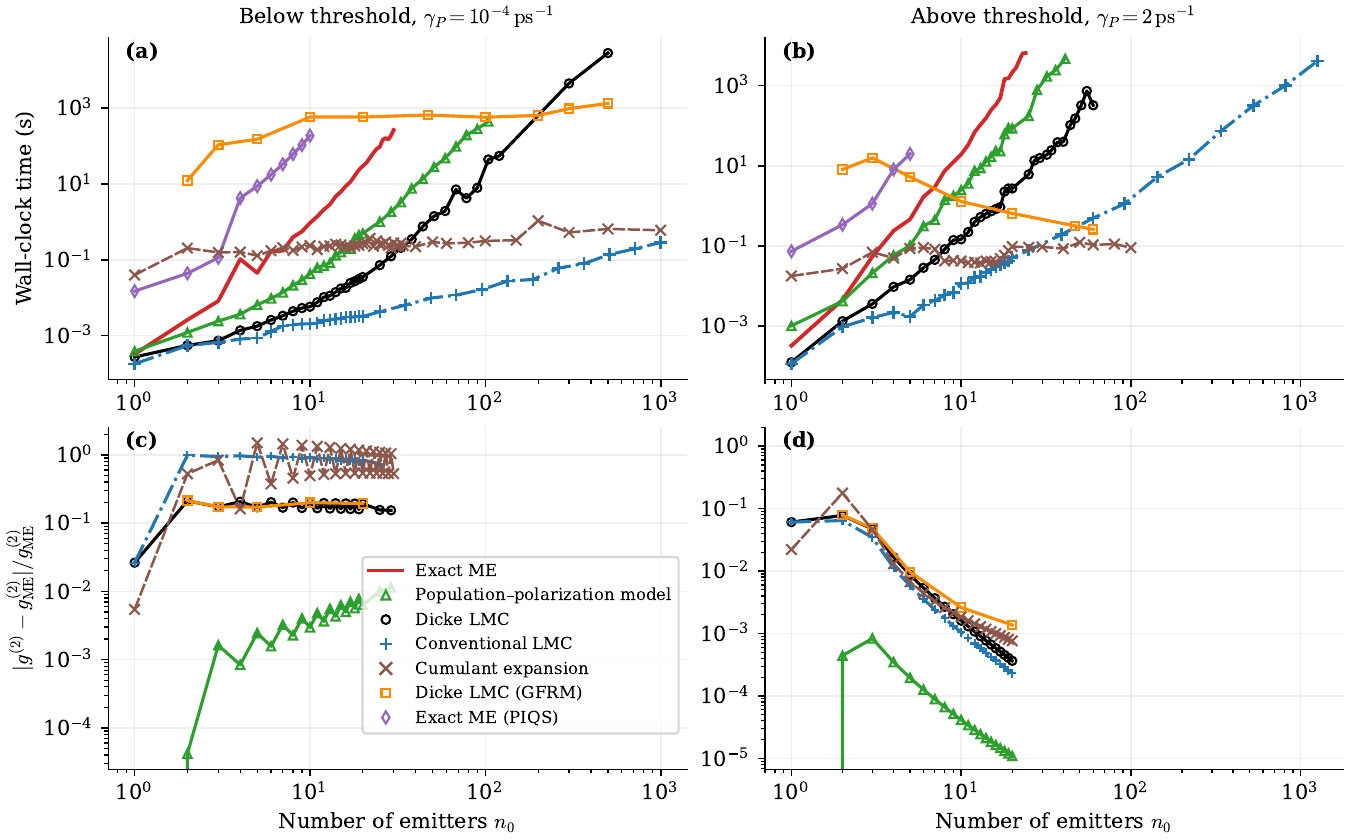}
   \caption{Computational cost and accuracy of the methods in Table~\ref{tab:methods} below threshold (left column, $\gamma_P = 10^{-4}\,\mathrm{ps}^{-1}$) and above threshold (right column, $\gamma_P = 2\,\mathrm{ps}^{-1}$). Parameters are the same as in Fig.~\ref{fig:bandwidth} except for the varied $n_0$ and $\gamma_P$. (a,b) Wall-clock time to obtain a converged steady state versus the number of emitters $n_0$. All solutions were computed sequentially on a single HPC node. For GFRM sampling of the Dicke LMC, the time is that needed to reach a relative accuracy of $10^{-3}$ in $g^{(2)}(0)$. For the fourth-order cumulant expansion, the time is per steady state and excludes the one-off symbolic derivation of the equations ($\approx 27$~s), which is independent of $n_0$ and of the rates. (c,d) Relative error in $g^{(2)}(0)$ with respect to the exact master equation, shown where the exact solution is feasible. The GFRM error is the model error of the Dicke LMC plus the $10^{-3}$ sampling tolerance.}
    \label{fig:benchmark}
\end{figure*}

In this subsection, we discuss the computational cost of the methods used in this paper. This highlights the different trade-offs between accuracy, speed, and scalability. The methods are compared in Fig.~\ref{fig:benchmark} below and above the laser threshold, where the relative error in $g^{(2)}(0)$ of each approximate method with respect to the exact ME is shown beneath the corresponding wall-clock times. All run times have been measured on an Intel Xeon Gold 6342 processor (2.8 GHz).

Below threshold, for small and moderate system sizes, the direct stationary solvers are the fastest. For $n_0 = 10$, the Dicke LMC solves for its $1.1\times10^3$ classical states in $6$~ms and the population--polarization model its $2.8\times10^3$ variables in $40$~ms, while the exact master equation takes on the order of a second, and the full PI Liouvillian of PIQS \cite{Shammah2018OpenInvariance}, constructed without the symmetry reduction, takes minutes at the same accuracy. The symmetry reduction of the master equation thus extends the reach of the exact solve to a few tens of emitters ($n_0 = 30$ costs $2.6\times10^2$~s, while memory becomes the limit near $n_0 \approx 34$). The population--polarization model can be solved up to $n_0 = 104$, and its cost lies an order of magnitude above that of the Dicke LMC but far below that of the exact ME, with excellent accuracy. It cannot, however, be sampled using GFRM, which limits its use for larger systems ($n_0>100$), where the large memory requirements to represent the total state space limit the usage of the approach.

GFRM sampling of the Dicke LMC costs of order $10^2-10^3$~s at the $10^{-3}$ accuracy target, essentially independently of the emitter number below threshold. It thus overtakes the exact solver around its memory limit and the stationary solve of the Dicke LMC at a few hundred emitters ($n_0\approx 300$). Above threshold, the cost of GFRM sampling even \emph{decreases} with $n_0$, and it becomes faster than the stationary solve beyond $n_0 \approx 20$. For many thousands of emitters, it is the only option.

The cumulant expansion costs $0.018$--$1.07$~s per steady state, nearly independently of $n_0$ both below and above threshold, since the number of equations does not grow with the emitter or photon number. This excludes the one-off symbolic derivation, which takes $\approx 27$~s at fourth order but grows steeply with the order (Appendix~\ref{sec:methods_cumulant}). Below threshold, it is cheaper than the direct solvers of the collective models beyond $n_0 \approx 30$, but it is the least accurate of the methods that capture collectivity. Above threshold, its accuracy is comparable to that of the Markov chains (an error of $2\times10^{-3}$ at $n_0 = 10$, decreasing with $n_0$), and it is cheaper than the Dicke LMC beyond $n_0 \approx 10$ and than the conventional LMC beyond $n_0 \approx 40$. Far above threshold, where the emitters are only weakly correlated, it is thus an efficient alternative.

The bottom row of Fig.~\ref{fig:benchmark} quantifies the improvement in numerical accuracy when higher polarization orders are taken into account. Below threshold, the Dicke LMC (and GFRM, which samples the same chain) has a nearly constant error of $\sim 20\%$ in $g^{(2)}(0)$ throughout the collective regime, while the population--polarization model is one to three orders of magnitude more accurate at every $n_0$. Its cost is, at the same time, closer to that of the Dicke LMC than to that of the exact solver. Above threshold, the Dicke LMC itself is accurate at the percent level (improving with $n_0$), and the population--polarization model improves on it by a further one to two orders of magnitude. We note that PIQS also offers a fast solver restricted to the diagonal of the Dicke-basis density matrix \cite{Shammah2018OpenInvariance}. It does not apply here, because the Tavis--Cummings Hamiltonian is not diagonal in the Dicke basis. The Dicke LMC can instead be viewed as the systematic construction of such a diagonal solver for the laser problem, with the polarization eliminated rather than ignored.

In summary, whenever the state space fits in memory, which in practice means up to several hundred emitters, the stationary solvers are the method of choice, with the population--polarization model as the more accurate option below threshold. Trajectory sampling is best reserved for very large systems, for time-dependent problems, or for observables along trajectories such as spectra \cite{Bundgaard-Nielsen2025SimpleNanolasers}. The cumulant expansion is a cheap option far above threshold or for strongly dephased emitters, but not in the collective regime below threshold.

\section{Conclusion}
\label{sec:conclusion}
We have investigated theoretical models for superradiant photon bunching in nanolasers. The key methodology is to sort the density matrix of the laser master equation by its polarization order, i.e., the number of photons involved in the coherence between the emitters and the light field. Because each polarization order only couples to its nearest neighbors, truncating at a given order and eliminating the highest retained order gives a systematic sequence of laser models, from the Dicke Laser Markov Chain to the exact master equation, in which each step adds one photon of coherence. Within this hierarchy, we have asked to what extent superradiant photon bunching in nanolasers can be understood as random jumps between Dicke states, and found three main conclusions.

First, superradiance can be predicted by a stochastic jump process with collectively modified rates. The Dicke LMC, a Markov chain over the photon number and the collective Dicke states, captures superthermal photon bunching and the even--odd dependence on the emitter number, neither of which are predicted by conventional laser rate equations. It recovers about $80\%$ of the bunching in the fully coherent limit, becomes accurate once dephasing exceeds the cavity loss rate, and, since it can be sampled at a cost independent of the number of emitters, it brings collective effects into stochastic laser modeling for thousands of emitters.

Second, the remaining bunching is an interference effect between successive emissions in the two-photon coherences. Keeping the optical polarization explicitly and eliminating the two-photon coherences gives the population--polarization model, which reduces the error in the bunching significantly, at a small fraction of the cost of the master equation. This model, however, cannot be formulated as a Markov chain with nonnegative rates. In this sense, most of the superradiant bunching is classical: once the collective Dicke states are identified, it follows from classical probability theory, as random jumps between bright and dark states with nonnegative rates, whereas the remainder requires coherence between the emitters and the light.

Third, regarding numerical methods, the symmetry-reduced master equation gives exact results across the whole lasing range for up to $24$ emitters, compared with $5$ for PIQS. Against this reference, the fourth-order cumulant expansion overestimates the bunching below threshold by about $50\%$ for ten emitters, and even more for odd emitter numbers, and it misses the even--odd effect entirely. Raising the order to $M = 8$ improves the accuracy only slowly, with an unstable steady state and a derivation time of $15$~h. The reason is that the emission is set by a small residue of nearly canceling emitter correlations, which a truncated moment expansion cannot resolve, while the Dicke models treat these correlations exactly. Far above threshold and for strong dephasing, where the emitters are nearly uncorrelated, the cumulant expansion is accurate and efficient.

Our benchmarks used a single set of cavity parameters, and mapping the validity of the models across coupling regimes is an important next step. Combining the Dicke LMC with the phase-resolved stochastic approach of Ref.~\cite{Bundgaard-Nielsen2025SimpleNanolasers} would give access to the linewidth of superradiant lasers \cite{Meiser2009ProspectsLaser,Bohnet2012ALaser,Norcia2016SuperradianceLaser}, and inhomogeneous emitter ensembles are a natural extension toward realistic quantum-dot lasers.

\begin{acknowledgments}
This work was supported by the Danish National Research Foundation through NanoPhoton - Center for Nanophotonics, Grant No. DNRF147.
\end{acknowledgments}

\section*{Code and data availability}
The code implementing all methods of Table~\ref{tab:methods}, the numerical data shown in the figures, and the scripts that reproduce the figures are openly available~\cite{Bundgaard-Nielsen2026SuperradiantLMCCode}.

\appendix

\section{Charge conservation of the laser Liouvillian}
\label{app:symmetry}
In this appendix, we prove the commutation relation $\comm{\mathcal{C}}{\mathcal{L}} = 0$ used in Sec.~\ref{sec:model}. Consider the global phase rotation generated by the total excitation number, $U_\phi = e^{i\phi N}$ with $N = a^\dagger a + J_z$. It acts on the basis states as
\begin{equation}
    U_\phi \ket{n_p, j, m} = e^{i\phi(n_p + m)} \ket{n_p, j, m} .
\end{equation}
Using $[N, a^\dagger] = a^\dagger$ and $[N, J_-] = -J_-$, we find
\begin{equation}
    [N, a^\dagger J_-] = [N, a^\dagger] J_- + a^\dagger [N, J_-] = a^\dagger J_- - a^\dagger J_- = 0 ,
\end{equation}
and likewise $[N, a J_+] = 0$, so $[H, N] = 0$.

The jump operators do not commute with $N$, but each transforms with a definite phase,
\begin{equation}
    U_\phi\, L\, U_\phi^\dagger = e^{i q_L \phi} L ,
    \qquad
    q_a = q_{\sigma_i} = -1, \quad q_{\sigma_i^\dagger} = +1, \quad q_{\sigma_i^\dagger\sigma_i} = 0 .
\end{equation}
In a Lindblad dissipator $\mathcal{D}_L(\rho) = L\rho L^\dagger - \tfrac12\{L^\dagger L, \rho\}$, the phase of $L$ cancels against the opposite phase of $L^\dagger$, so
\begin{equation}
    \mathcal{D}_L\!\left(U_\phi\, \rho\, U_\phi^\dagger\right) = U_\phi\, \mathcal{D}_L(\rho)\, U_\phi^\dagger
\end{equation}
for every operator $L$. Together with $[H,N]=0$, this establishes the covariance $\mathcal{L}[U_\phi\rho U_\phi^\dagger] = U_\phi\mathcal{L}[\rho]U_\phi^\dagger$ for all $\phi$, i.e., $[\mathcal{L}, \mathcal{C}] = 0$ with $\mathcal{C}\rho = [N, \rho]$ \cite{Albert2014SymmetriesEquations}.

\section{Projection and closure steps}
\label{app:derivation}
In this appendix, we collect the technical steps behind Secs.~\ref{sec:hierarchy} and \ref{sec:elimination}, following the route established for the conventional LMC in Ref.~\cite{Bundgaard-Nielsen2026ModelingChains}.

\subsection{Markov approximation}
Formally integrating Eq.~\eqref{eq:block_q} with the truncated generator $\Lb{1}{1}$ gives
\begin{equation}
    Q(t) = e^{\Lb{1}{1} t}\,Q(0) + \int_0^t ds\; e^{\Lb{1}{1}(t-s)}\, \Lb{1}{0}\,P(s) .
\end{equation}
For a diagonal initial state, $Q(0) = 0$. The Markov approximation replaces $P(s) \to P(t)$ inside the integral and extends the upper limit to infinity, which is justified when the populations vary slowly on the decay timescale of $\Lb{1}{1}$; this yields Eq.~\eqref{eq:adiabatic}. Since the matrix elements of $\mathcal{L}_H$ between the relevant states carry the collective factors, $\bra{n_p+1, j, m-1} H \ket{n_p, j, m} = g\sqrt{n_p+1}\sqrt{C_-(j,m)} \equiv G_-$, the parameter controlling this approximation is $G_-/\gamma_\perp$ [Eq.~\eqref{eq:validity}], not $g/\gamma_\perp$. The same steps applied to the $B=2$ system, with $R$ in place of $Q$ and $\Lb{2}{2}$ in place of $\Lb{1}{1}$, give the elimination used in Sec.~\ref{sec:secondorder}; at stationarity the elimination is exact given the closure of $\Lb{2}{2}$, no Markov approximation being required for the retained dynamical $Q$.

\subsection{Collective eigenoperators of the dissipation}
\label{app:eigenoperators}
Applying the adjoint dissipators to $O_- = a^\dagger J_-$ term by term,
\begin{equation}
\begin{aligned}
    \gamma_c \mathcal{D}^\dagger_a (O_-) &= -\tfrac{\gamma_c}{2}\, O_-, \\
    \textstyle\sum_i \gamma_A \mathcal{D}^\dagger_{\sigma_i}(O_-) &= -\tfrac{\gamma_A}{2}\, O_-, \\
    \textstyle\sum_i \gamma_P \mathcal{D}^\dagger_{\sigma_i^\dagger}(O_-) &= -\tfrac{\gamma_P}{2}\, O_-, \\
    \textstyle\sum_i \gamma_D \mathcal{D}^\dagger_{\sigma_i^\dagger\sigma_i}(O_-) &= -\tfrac{\gamma_D}{2}\, O_-,
\end{aligned}
\label{eq:app_eigen_terms}
\end{equation}
establishes Eq.~\eqref{eq:eigenoperator}: $O_-$ (and likewise $O_+ = a J_+$) is a left eigenoperator of $\mathcal{L}_D$ with eigenvalue $-\gamma_\perp/2$. Consequently, for any $X$ in the polarization sector,
\begin{equation}
    \Tr{O_-\, \Lb{1}{1}^{-1} X} = -\frac{2}{\gamma_\perp} \Tr{O_-\, X},
\end{equation}
i.e., the closure in Eq.~\eqref{eq:scalar_closure} is exact at the level of the collective polarization trace. What it neglects is the internal redistribution among the individual coherence elements. To see what this means concretely, consider the photon part of a polarization element, $q_{n} = \bra{n+1}\rho\ket{n}$, under cavity decay alone:
\begin{equation}
    \dot q_n = \gamma_c \left[ \sqrt{(n+1)(n+2)}\, q_{n+1} - \left(n + \tfrac{1}{2}\right) q_n \right],
    \label{eq:app_qn}
\end{equation}
so the individual elements are coupled along the photon ladder. Yet for the polarization expectation value $\langle a \rangle = \sum_n \sqrt{n+1}\, q_n$, the couplings conspire (shift the summation index in the first term) to give exactly $\frac{d}{dt}\langle a\rangle = -\tfrac{\gamma_c}{2}\langle a\rangle$. Replacing Eq.~\eqref{eq:app_qn} by $\dot q_n = -\tfrac{\gamma_c}{2} q_n$, which is what the closure does, preserves this expectation value while discarding the element-wise redistribution, in close analogy to the factorization step of a cumulant expansion \cite{Bundgaard-Nielsen2026ModelingChains}.

The two-photon analogue follows by the same term-by-term computation applied to $O_-^{(2)} = (a^\dagger)^2 J_-^2$. For the cavity, using $a (a^\dagger)^2 = (a^\dagger)^2 a + 2a^\dagger$,
\begin{equation}
    \mathcal{D}^\dagger_a\!\left[(a^\dagger)^2\right] = a^\dagger (a^\dagger)^2 a - \tfrac12\{a^\dagger a, (a^\dagger)^2\} = -\,(a^\dagger)^2 ,
\end{equation}
i.e., twice the single-photon eigenvalue. For the local dissipators, we expand $J_-^2 = \sum_{i \ne k}\sigma_i\sigma_k$ (the $i=k$ terms vanish). Each site $l \in \{i, k\}$ then contributes $-\gamma_\mu/2$ exactly as in Eq.~\eqref{eq:app_eigen_terms}, while all other sites contribute zero, so each dissipator contributes $-\gamma_\mu$ in total. Summing,
\begin{equation}
    \mathcal{L}_D^\dagger\!\left[(a^\dagger)^2 J_-^2\right] = -\gamma_\perp\, (a^\dagger)^2 J_-^2 ,
\end{equation}
which is Eq.~\eqref{eq:eigenoperator2}; the pattern continues as $-k\gamma_\perp/2$ for $(a^\dagger)^k J_-^k$.

\section{Dissipative branching rates}
\label{app:rates}
The action of sums of local operators on Dicke density-matrix elements follows Refs.~\cite{Chase2008CollectiveParticles,Baragiola2010CollectiveSystems,Shammah2018OpenInvariance},
\begin{equation}
\begin{aligned}
    \sum_{i=1}^{n_0} J_{q,i}& \ketbra{j,m}{j,m'} J_{r,i}^\dagger
    = a^{n_0}_{qr}\, \ketbra{j,m{+}\tilde q}{j,m'{+}\tilde r} \\
    &+ b^{n_0}_{qr}\, \ketbra{j{-}1,m{+}\tilde q}{j{-}1,m'{+}\tilde r} \\
    &+ c^{n_0}_{qr}\, \ketbra{j{+}1,m{+}\tilde q}{j{+}1,m'{+}\tilde r},
\end{aligned}
\label{eq:app_branching}
\end{equation}
with $q, r \in \{+,-,z\}$ for $J_{+,i} = \sigma_i^\dagger$, $J_{-,i} = \sigma_i$, $J_{z,i} = \sigma_i^\dagger\sigma_i$ and $\tilde q, \tilde r \in \{+1,-1,0\}$; the arguments $(j,m,m')$ of the coefficients $a$, $b$, $c$ are suppressed. The coefficients factorize into $j$-dependent prefactors and Dicke matrix elements,
\begin{equation}
\begin{aligned}
    a^{n_0}_{qr}(j,m,m') &= A^{j,m}_q A^{j,m'}_r\, \frac{S+1}{2j(j+1)}, \\
    b^{n_0}_{qr}(j,m,m') &= B^{j,m}_q B^{j,m'}_r\, \frac{S+j+1}{2j(2j+1)}, \\
    c^{n_0}_{qr}(j,m,m') &= D^{j,m}_q D^{j,m'}_r\, \frac{S-j}{2(j+1)(2j+1)},
\end{aligned}
\label{eq:app_abc}
\end{equation}
with $S = n_0/2$ and
\begin{equation}
\begin{aligned}
    A^{j,m}_\pm &= \sqrt{(j \mp m)(j \pm m + 1)}, \\
    B^{j,m}_\pm &= \pm\sqrt{(j \mp m)(j \mp m - 1)}, \\
    D^{j,m}_\pm &= \mp\sqrt{(j \pm m + 1)(j \pm m + 2)}, \\
    A^{j,m}_z &= m, \qquad B^{j,m}_z = \sqrt{(j+m)(j-m)}, \\
    D^{j,m}_z &= \sqrt{(j{+}m{+}1)(j{-}m{+}1)} .
\end{aligned}
\end{equation}
For the population projection of Appendix~\ref{app:derivation} only the diagonal case $m' = m$, $r = q$ is needed, and only the squared matrix elements enter,
\begin{equation}
\begin{aligned}
    |A^{j,m}_\pm|^2 &= (j \mp m)(j \pm m + 1), \\
    |B^{j,m}_\pm|^2 &= (j \mp m)(j \mp m - 1), \\
    |D^{j,m}_\pm|^2 &= (j \pm m + 1)(j \pm m + 2), \\
    |A^{j,m}_z|^2 &= m^2, \qquad |B^{j,m}_z|^2 = j^2 - m^2, \\
    |D^{j,m}_z|^2 &= (j{+}1)^2 - m^2 ,
\end{aligned}
\end{equation}
so the signs of $B^{j,m}_\pm$ and $D^{j,m}_\pm$ never enter the transition rates. Combining Eqs.~\eqref{eq:app_branching}--\eqref{eq:app_abc} with the prefactors $\eta_{0,\pm}(j)$ of Eq.~\eqref{eq:eta} yields the pump, decay, and dephasing rows of Table~\ref{tab:events_dicke}. As an example, the background-decay dissipator $\sum_i \gamma_A \mathcal{D}_{\sigma_i}$ has $q = r = -$ and branches into the three events
\begin{equation}
    (j,m) \to (j, m{-}1),\quad (j{-}1, m{-}1),\quad (j{+}1, m{-}1),
\end{equation}
with rates $\gamma_A\eta_0(j)|A^{j,m}_-|^2$, $\gamma_A\eta_-(j)|B^{j,m}_-|^2$, and $\gamma_A\eta_+(j)|D^{j,m}_-|^2$, which sum to $\gamma_A(S+m) = \gamma_A n_e$ as required by trace preservation. The same-$j$ part of the pure-dephasing dissipator, $\gamma_D \eta_0(j)|A^{j,m}_z|^2$, cancels exactly against the anticommutator in the projected population equation, since it does not change the resolved state; only the $j \to j\pm1$ branches of Table~\ref{tab:events_dicke} survive. Note that defining pure dephasing with the jump operator $\sigma_z$ instead of $\sigma^\dagger\sigma$ rescales the rate by a factor of four, $\gamma_z = \gamma_D/4$. The general off-diagonal case $m \ne m'$ of Eq.~\eqref{eq:app_branching} is what enters the dissipative blocks $\Lb{b}{b}$ at $b \ge 1$, used by the population--polarization model and the reference solver.

\section{Reduction of the Dicke LMC to the conventional LMC}
\label{app:lmc_limit}
In this appendix, we make precise the statement of Sec.~\ref{sec:lmc} that the conventional LMC over $(n_p, n_e)$ is the $j$-agnostic limit of the Dicke LMC. The natural link between the two state spaces is the marginal probability at fixed number of excited emitters $n_e = m + n_0/2$,
\begin{equation}
    \bar{P}(n_p, n_e) = \sum_j P(n_p, j, m)\Big|_{m\, =\, n_e - n_0/2} .
    \label{eq:marginal}
\end{equation}
Summing the Dicke LMC master equation [Eq.~\eqref{eq:cme} with Table~\ref{tab:events_dicke}] over $j$ shows, event by event, which rates close on the marginal and which do not.

\emph{Cavity decay} does not involve the emitter state and closes trivially.

\emph{Local decay and pump} change $j$ by $0, \pm1$, but always change $n_e$ by the same amount ($\mp 1$). Their three branch rates sum to the uncorrelated-emitter totals independently of $j$ (Appendix~\ref{app:rates}),
\begin{equation}
\begin{aligned}
    \gamma_A \left[ \eta_0 |A^{j,m}_-|^2 + \eta_- |B^{j,m}_-|^2 + \eta_+ |D^{j,m}_-|^2 \right] &= \gamma_A\, n_e , \\
    \gamma_P \left[ \eta_0 |A^{j,m}_+|^2 + \eta_- |B^{j,m}_+|^2 + \eta_+ |D^{j,m}_+|^2 \right] &= \gamma_P (n_0 - n_e) ,
\end{aligned}
\label{eq:branch_sums}
\end{equation}
so these events appear in the marginal dynamics exactly as in the conventional LMC.

\emph{Pure dephasing} changes only $j$, at fixed $n_p$ and $n_e$. On the marginal in Eq.~\eqref{eq:marginal}, the gain of each branch cancels against the loss of another, and dephasing disappears from the dynamics altogether: it is not an event type of the conventional LMC.

\emph{The radiative events} are the only ones that do not close: their rates carry the Dicke factors $C_\pm(j,m)$, which depend on $j$. The conventional LMC follows from the single replacement
\begin{equation}
    C_-(j,m) \to n_e , \qquad C_+(j,m) \to n_0 - n_e ,
    \label{eq:mean_field_replacement}
\end{equation}
which, by Eq.~\eqref{eq:cminus_pairwise}, amounts to neglecting the pairwise interemitter coherences $\langle \sigma_i^\dagger \sigma_k \rangle$. This step, and only this step, distinguishes the two chains: all collective physics of the Dicke LMC resides in the radiative rates. Table~\ref{tab:lmc_limit} summarizes the correspondence.

\begin{table}[!ht]
\caption{Event-by-event correspondence between the Dicke LMC (Table~\ref{tab:events_dicke}) and the conventional LMC. The local decay and pump rows of the Dicke LMC comprise three branches each, whose rates sum to the conventional rate [Eq.~\eqref{eq:branch_sums}]; only the radiative rows involve an approximation, Eq.~\eqref{eq:mean_field_replacement}.
\label{tab:lmc_limit}}
\begin{ruledtabular}
\begin{tabular}{lcc}
\textbf{Event} & \textbf{Dicke LMC} & \textbf{Conventional LMC} \\
\colrule
Collective emission & $\gamma_r (n_p{+}1)\, C_-(j,m)$ & $\gamma_r (n_p{+}1)\, n_e$ \\
Collective absorption & $\gamma_r n_p\, C_+(j,m)$ & $\gamma_r n_p (n_0{-}n_e)$ \\
Cavity decay & $\gamma_c n_p$ & $\gamma_c n_p$ \\
Decay ($j \to j, j{\pm}1$) & three branches & $\gamma_A n_e$ \\
Pump ($j \to j, j{\pm}1$) & three branches & $\gamma_P (n_0{-}n_e)$ \\
Dephasing ($j \to j{\pm}1$) & two branches & absent \\
\end{tabular}
\end{ruledtabular}
\end{table}

\section{Numerical methods}
\label{app:numerical_details}
In this appendix, we discuss the numerical approaches used to solve the Dicke LMC and the population--polarization model, namely direct stationary solves of the sparse generators and trajectory sampling with Gillespie's first reaction method. We also summarize how the cumulant expansion is solved.

\subsection{Direct stationary solution of the LMC \label{sec:direct}}
For the Dicke LMC in Eq.~\eqref{eq:diagonal_eq_final}, the population vector $P$ evolves linearly according to $\dot{P} = M P$, with a sparse generator $M$ constructed according to Eq.~\eqref{eq:cme}. For steady-state observables, we can solve $M P = 0$ directly. To construct $M$, the photon space is truncated at a cutoff $n_{\max}$, and the cutoff is increased until the stationary probability at $n_{\max}$ is below $10^{-10}$. The number of states is $(n_{\max}+1) \times d_{n_0}$, where $d_{n_0}$ counts the $(j,m)$ pairs, $d_{n_0} = (S+1)^2$ for even and $(S+\tfrac12)(S+\tfrac32)$ for odd $n_0$ \cite{Shammah2018OpenInvariance}; i.e., the state space grows only quadratically with the emitter number at fixed photon cutoff. The stationary vector is obtained by shifted inverse iteration: we factorize $\sigma I - M$ once, with $\sigma$ many orders of magnitude below the smallest physical rate, and iterate $P \leftarrow (\sigma I - M)^{-1} P$ with normalization until $\|M P\|_\infty < 10^{-11}$. Because $M$ is sparse and the shift preserves its sparsity, this scales to hundreds of thousands of states. From $P$, all stationary observables follow directly, e.g., $\langle n_p \rangle = \sum_{\mathbf{x}} n_p\, P(\mathbf{x})$ and $g^{(2)}(0) = \langle n_p(n_p-1)\rangle / \langle n_p\rangle^2$.

Delayed correlations are equally accessible. The detection of a photon at $\tau = 0$ removes one photon and reweights the stationary distribution by $n_p$, giving the conditioned distribution
\begin{equation}
    P_c(n_p{-}1, j, m) = \frac{n_p\, P_{\mathrm{ss}}(n_p,j,m)}{\langle n_p\rangle} ,
    \label{eq:conditioned}
\end{equation}
which is then propagated with the generator and read out with the photon number,
\begin{equation}
    g^{(2)}(\tau) = \frac{1}{\langle n_p\rangle}\sum_{\mathbf{x}} n_p\, \left[e^{M\tau}\, P_{c}\right](\mathbf{x}) .
    \label{eq:g2tau}
\end{equation}
The matrix exponential is never formed explicitly. Instead, $e^{M\tau} P_c$ is advanced with the Crank--Nicolson scheme,
\begin{equation}
    \left[I - \tfrac{\Delta t}{2} M\right] P(t + \Delta t)
    = \left[I + \tfrac{\Delta t}{2} M\right] P(t) ,
\end{equation}
on a piecewise time grid whose step size grows with $\tau$, so that each segment of the grid reuses a single sparse LU factorization of $I - \tfrac{\Delta t}{2}M$. The same construction, with $M$ replaced by the corresponding generator and the classical detection map~\eqref{eq:conditioned} by $a \rho\, a^\dagger / \langle n_p \rangle$ (the quantum regression theorem), is used for the population--polarization model and for the exact ME reference in Sec.~\ref{sec:results_g2tau}.

\subsection{Trajectory sampling of the Dicke LMC \label{sec:gfrm}}
Alternatively, Eq.~\eqref{eq:cme} can be sampled with Gillespie's first reaction method (GFRM) \cite{Gillespie1976AReactions}: from the current state, a waiting time is drawn for each event of Table~\ref{tab:events_dicke} and the earliest event is executed. Time averages over long trajectories converge to the stationary distribution \cite{Gillespie1992AEquation,Gillespie2007StochasticKinetics}. This is how the LMC has previously been sampled \cite{Roy-Choudhury2009QuantumLasers,Roy-Choudhury2010QuantumDiodes,Lebreton2013StochasticallyNanolasers,Puccioni2015StochasticLasing,Mork2018RateEmitters,Takemura2019LasingLasers,Andre2020,Lippi2021AmplifiedNanolasers,Bundgaard-Nielsen2023StochasticNanolaser,Puccioni2024StatisticsNanolasers,Bundgaard-Nielsen2025SimpleNanolasers,Bundgaard-Nielsen2026ModelingChains}. The cost per event is independent of both $n_0$ and $n_p$, so sampling remains feasible for arbitrarily large systems, which makes it well suited for systems with large populations. In certain regimes, tau-leaping \cite{Gillespie2001ApproximateSystems}, which advances many events per step, can be numerically more efficient, and for very large populations, it connects smoothly to the Langevin rate-equation limit, which provides an even more efficient sampling \cite{Bundgaard-Nielsen2026ModelingChains}.

Compared with the direct stationary solution of Appendix~\ref{sec:direct}, the drawback of GFRM is that many samples are needed to reduce the statistical error of $g^{(2)}(0)$, which is often more costly numerically, as we discuss in Sec.~\ref{sec:results_cost}.

\subsection{Population--polarization solver}
\label{sec:methods_resum}
The population--polarization model in Eq.~\eqref{eq:resum_model} is solved in the sparse augmented form
\begin{equation}
    \begin{pmatrix} \Lb{0}{0} & \Lb{0}{1} \\ \Lb{1}{0} & F \end{pmatrix}
    \begin{pmatrix} P \\ Q \end{pmatrix} = 0 ,
    \qquad \textstyle\sum_{\mathbf{x}} P_{\mathbf{x}} = 1 ,
    \label{eq:augmented}
\end{equation}
by the same shifted inverse iteration as the Dicke LMC. Delayed correlations follow from the quantum regression theorem applied in the augmented space: the detection superoperator $a\,\rho\, a^\dagger$ preserves both the charge and the polarization order, so $g^{(2)}(\tau)$ is obtained by propagating with the generator of Eq.~\eqref{eq:augmented} exactly as in Eq.~\eqref{eq:g2tau}. The stationary populations $P$ of the population--polarization model can acquire negative values at the $10^{-6}$ level, inherited from the signed effective weights discussed in Sec.~\ref{sec:secondorder}. They are irrelevant for the low-order moments reported here, but they prevent sampling the model as a jump process, so the trajectory picture is only available at first order.

\subsection{Cumulant expansion}
\label{sec:methods_cumulant}
The fourth-order cumulant equations are derived once with QuantumCumulants.jl \cite{Plankensteiner2022quantumcumulantsjl}, using indexed emitter operators and its scaling for identical emitters; the derivation takes $\approx 27$~s and is independent of $n_0$ and of the rates, so it is reused for all parameter points. Moments with nonzero charge are dropped, which is exact for a phase-invariant state, leaving $18$ equations. Moments involving more distinct emitters than $n_0$ do not exist and are set to zero; this is exact and is needed to keep the integration stable for $n_0 \le 2$. We integrate the equations from the physical initial state, the cavity vacuum with all emitters in $\ket{g}$, with a stiff Rosenbrock solver, and accept the result as a steady state only once the scaled residual of the right-hand side is below $10^{-8}$. For $g^{(2)}(\tau)$, the initial moments of the conditional state $a\rho a^\dagger/\expval{n_p}$ are obtained from the stationary moments, with moments beyond fourth order factorized by the same closure, and are then propagated with the same equations.

The higher orders in Fig.~\ref{fig:bandwidth} use the same derivation, with $4$, $8$, $18$, $28$, $48$, $68$, and $103$ equations for $M = 2$--$8$. The derivation time grows steeply with the order, since the closure factorizes every moment of order $M+1$ over all of its set partitions: on a single core, it takes $2$~min up to $M = 6$, $29$~min at $M = 7$, and $15$~h at $M = 8$, where a subsequent solution is found after $\sim 30 \min$. We observe that the resulting equations of motion are not guaranteed to have stable fixed points. I.e., their Jacobian has an eigenvalue with a positive real part; thus, time integration does not find this solution. This occurs at $M = 4$ near threshold and for $n_0 \gtrsim 150$ below threshold, and at the test case for all $M \ge 5$. There we solve for the fixed point directly with a Newton method (solving $\mathbf{F}(\mathbf{u}) = 0$ by iterating $\mathbf{u} \to \mathbf{u} - \mathbf{J}^{-1}\mathbf{F}(\mathbf{u})$, where $\dot{\mathbf{u}} = \mathbf{F}(\mathbf{u})$ denotes the truncated equations for the vector of moments $\mathbf{u}$ and $\mathbf{J} = \partial\mathbf{F}/\partial\mathbf{u}$), with the Jacobian $\mathbf{J}$ obtained by automatic differentiation.

\section{Symmetry-reduced solvers}
\label{app:reference}
The reference solver implements the hierarchy of Sec.~\ref{sec:hierarchy} directly. We enumerate all density-matrix elements $\rho_{n_p j m;\, n_p' j m'}$ satisfying the charge constraint $n_p + m = n_p' + m'$ [Eq.~\eqref{eq:u1}] and the truncation $|n_p - n_p'| \le B$, up to a photon cutoff $n_{\max}$, and assemble the sparse Liouvillian in this basis: the Hamiltonian contributes first-order couplings between neighboring polarization orders, the cavity dissipator couples along the photon ladder at fixed offsets, and the local dissipators enter through the branching identities of Eq.~\eqref{eq:app_branching}, including the off-diagonal case $m \ne m'$. The stationary element vector is obtained by shifted inverse iteration as in Appendix~\ref{sec:direct}.

Convergence is controlled by two parameters. The photon cutoff is verified through the stationary population at the cutoff edge (required $<10^{-10}$) and enlarged if necessary, exactly as for the classical chain. The truncation order is increased through $B = 1, 2, 4, 6, \dots$ until $\langle n_p\rangle$ and $g^{(2)}(0)$ change by less than $10^{-6}$ between consecutive solves. If the criterion is not met, the sequence terminates at $B = \min(n_{\max}, n_0)$, where the solution coincides with the PI master equation by the symmetry argument of Sec.~\ref{sec:hierarchy}. For the test case of Fig.~\ref{fig:bandwidth} ($n_0 = 10$, $\gamma_P = 10^{-4}\,\mathrm{ps}^{-1}$), this procedure reproduces the steady state computed with PIQS \cite{Shammah2018OpenInvariance} (direct sparse LU method) to $2\times10^{-8}$ in $g^{(2)}(0)$ and to better than $10^{-8}$ in $\langle n_p \rangle$.


\begin{thebibliography}{58}%
\makeatletter
\providecommand \@ifxundefined [1]{%
 \@ifx{#1\undefined}
}%
\providecommand \@ifnum [1]{%
 \ifnum #1\expandafter \@firstoftwo
 \else \expandafter \@secondoftwo
 \fi
}%
\providecommand \@ifx [1]{%
 \ifx #1\expandafter \@firstoftwo
 \else \expandafter \@secondoftwo
 \fi
}%
\providecommand \natexlab [1]{#1}%
\providecommand \enquote  [1]{``#1''}%
\providecommand \bibnamefont  [1]{#1}%
\providecommand \bibfnamefont [1]{#1}%
\providecommand \citenamefont [1]{#1}%
\providecommand \href@noop [0]{\@secondoftwo}%
\providecommand \href [0]{\begingroup \@sanitize@url \@href}%
\providecommand \@href[1]{\@@startlink{#1}\@@href}%
\providecommand \@@href[1]{\endgroup#1\@@endlink}%
\providecommand \@sanitize@url [0]{\catcode `\\12\catcode `\$12\catcode `\&12\catcode `\#12\catcode `\^12\catcode `\_12\catcode `\%12\relax}%
\providecommand \@@startlink[1]{}%
\providecommand \@@endlink[0]{}%
\providecommand \url  [0]{\begingroup\@sanitize@url \@url }%
\providecommand \@url [1]{\endgroup\@href {#1}{\urlprefix }}%
\providecommand \urlprefix  [0]{URL }%
\providecommand \Eprint [0]{\href }%
\providecommand \doibase [0]{https://doi.org/}%
\providecommand \selectlanguage [0]{\@gobble}%
\providecommand \bibinfo  [0]{\@secondoftwo}%
\providecommand \bibfield  [0]{\@secondoftwo}%
\providecommand \translation [1]{[#1]}%
\providecommand \BibitemOpen [0]{}%
\providecommand \bibitemStop [0]{}%
\providecommand \bibitemNoStop [0]{.\EOS\space}%
\providecommand \EOS [0]{\spacefactor3000\relax}%
\providecommand \BibitemShut  [1]{\csname bibitem#1\endcsname}%
\let\auto@bib@innerbib\@empty
\bibitem [{\citenamefont {Ma}\ and\ \citenamefont {Oulton}(2019)}]{Ma2019ApplicationsNanolasers}%
  \BibitemOpen
  \bibfield  {author} {\bibinfo {author} {\bibfnamefont {R.-M.}\ \bibnamefont {Ma}}\ and\ \bibinfo {author} {\bibfnamefont {R.~F.}\ \bibnamefont {Oulton}},\ }\href {https://doi.org/10.1038/s41565-018-0320-y} {\bibfield  {journal} {\bibinfo  {journal} {Nature Nanotechnology}\ }\textbf {\bibinfo {volume} {14}},\ \bibinfo {pages} {12} (\bibinfo {year} {2019})}\BibitemShut {NoStop}%
\bibitem [{\citenamefont {Dimopoulos}\ \emph {et~al.}(2022)\citenamefont {Dimopoulos}, \citenamefont {Sakanas}, \citenamefont {Marchevsky}, \citenamefont {Xiong}, \citenamefont {Yu}, \citenamefont {Semenova}, \citenamefont {M{\o}rk},\ and\ \citenamefont {Yvind}}]{Dimopoulos2022Electrically-DrivenThreshold}%
  \BibitemOpen
  \bibfield  {author} {\bibinfo {author} {\bibfnamefont {E.}~\bibnamefont {Dimopoulos}}, \bibinfo {author} {\bibfnamefont {A.}~\bibnamefont {Sakanas}}, \bibinfo {author} {\bibfnamefont {A.}~\bibnamefont {Marchevsky}}, \bibinfo {author} {\bibfnamefont {M.}~\bibnamefont {Xiong}}, \bibinfo {author} {\bibfnamefont {Y.}~\bibnamefont {Yu}}, \bibinfo {author} {\bibfnamefont {E.}~\bibnamefont {Semenova}}, \bibinfo {author} {\bibfnamefont {J.}~\bibnamefont {M{\o}rk}},\ and\ \bibinfo {author} {\bibfnamefont {K.}~\bibnamefont {Yvind}},\ }\href {https://doi.org/10.1002/lpor.202200109} {\bibfield  {journal} {\bibinfo  {journal} {Laser and Photonics Reviews}\ }\textbf {\bibinfo {volume} {16}},\ \bibinfo {pages} {2200109} (\bibinfo {year} {2022})}\BibitemShut {NoStop}%
\bibitem [{\citenamefont {Rice}\ and\ \citenamefont {Carmichael}(1994)}]{Rice1994PhotonAnalogy}%
  \BibitemOpen
  \bibfield  {author} {\bibinfo {author} {\bibfnamefont {P.~R.}\ \bibnamefont {Rice}}\ and\ \bibinfo {author} {\bibfnamefont {H.~J.}\ \bibnamefont {Carmichael}},\ }\href {https://doi.org/10.1103/PhysRevA.50.4318} {\bibfield  {journal} {\bibinfo  {journal} {Physical Review A}\ }\textbf {\bibinfo {volume} {50}},\ \bibinfo {pages} {4318} (\bibinfo {year} {1994})}\BibitemShut {NoStop}%
\bibitem [{\citenamefont {Saldutti}\ \emph {et~al.}(2024)\citenamefont {Saldutti}, \citenamefont {Yu},\ and\ \citenamefont {M{\o}rk}}]{Saldutti2024TheNanolasers}%
  \BibitemOpen
  \bibfield  {author} {\bibinfo {author} {\bibfnamefont {M.}~\bibnamefont {Saldutti}}, \bibinfo {author} {\bibfnamefont {Y.}~\bibnamefont {Yu}},\ and\ \bibinfo {author} {\bibfnamefont {J.}~\bibnamefont {M{\o}rk}},\ }\href {https://doi.org/10.1002/lpor.202300840} {\bibfield  {journal} {\bibinfo  {journal} {Laser \& Photonics Reviews}\ }\textbf {\bibinfo {volume} {18}},\ \bibinfo {pages} {2300840} (\bibinfo {year} {2024})}\BibitemShut {NoStop}%
\bibitem [{\citenamefont {M{\o}rk}\ and\ \citenamefont {Yvind}(2020)}]{Mork2020SqueezingConfinement}%
  \BibitemOpen
  \bibfield  {author} {\bibinfo {author} {\bibfnamefont {J.}~\bibnamefont {M{\o}rk}}\ and\ \bibinfo {author} {\bibfnamefont {K.}~\bibnamefont {Yvind}},\ }\href {https://doi.org/10.1364/OPTICA.402190} {\bibfield  {journal} {\bibinfo  {journal} {Optica}\ }\textbf {\bibinfo {volume} {7}},\ \bibinfo {pages} {1641} (\bibinfo {year} {2020})}\BibitemShut {NoStop}%
\bibitem [{\citenamefont {Dicke}(1954)}]{Dicke1954CoherenceProcesses}%
  \BibitemOpen
  \bibfield  {author} {\bibinfo {author} {\bibfnamefont {R.~H.}\ \bibnamefont {Dicke}},\ }\href {https://doi.org/10.1103/PhysRev.93.99} {\bibfield  {journal} {\bibinfo  {journal} {Physical Review}\ }\textbf {\bibinfo {volume} {93}},\ \bibinfo {pages} {99} (\bibinfo {year} {1954})}\BibitemShut {NoStop}%
\bibitem [{\citenamefont {Gross}\ and\ \citenamefont {Haroche}(1982)}]{Gross1982SuperradianceEmission}%
  \BibitemOpen
  \bibfield  {author} {\bibinfo {author} {\bibfnamefont {M.}~\bibnamefont {Gross}}\ and\ \bibinfo {author} {\bibfnamefont {S.}~\bibnamefont {Haroche}},\ }\href {https://doi.org/10.1016/0370-1573(82)90102-8} {\bibfield  {journal} {\bibinfo  {journal} {Physics Reports}\ }\textbf {\bibinfo {volume} {93}},\ \bibinfo {pages} {301} (\bibinfo {year} {1982})}\BibitemShut {NoStop}%
\bibitem [{\citenamefont {Temnov}\ and\ \citenamefont {Woggon}(2005)}]{Temnov2005SuperradianceCavity}%
  \BibitemOpen
  \bibfield  {author} {\bibinfo {author} {\bibfnamefont {V.~V.}\ \bibnamefont {Temnov}}\ and\ \bibinfo {author} {\bibfnamefont {U.}~\bibnamefont {Woggon}},\ }\href {https://doi.org/10.1103/PhysRevLett.95.243602} {\bibfield  {journal} {\bibinfo  {journal} {Physical Review Letters}\ }\textbf {\bibinfo {volume} {95}},\ \bibinfo {pages} {243602} (\bibinfo {year} {2005})}\BibitemShut {NoStop}%
\bibitem [{\citenamefont {Auff{\`e}ves}\ \emph {et~al.}(2011)\citenamefont {Auff{\`e}ves}, \citenamefont {Gerace}, \citenamefont {Portolan}, \citenamefont {Drezet},\ and\ \citenamefont {Fran{\c{c}}a~Santos}}]{Auffeves2011FewIndividualization}%
  \BibitemOpen
  \bibfield  {author} {\bibinfo {author} {\bibfnamefont {A.}~\bibnamefont {Auff{\`e}ves}}, \bibinfo {author} {\bibfnamefont {D.}~\bibnamefont {Gerace}}, \bibinfo {author} {\bibfnamefont {S.}~\bibnamefont {Portolan}}, \bibinfo {author} {\bibfnamefont {A.}~\bibnamefont {Drezet}},\ and\ \bibinfo {author} {\bibfnamefont {M.}~\bibnamefont {Fran{\c{c}}a~Santos}},\ }\href {https://doi.org/10.1088/1367-2630/13/9/093020} {\bibfield  {journal} {\bibinfo  {journal} {New Journal of Physics}\ }\textbf {\bibinfo {volume} {13}},\ \bibinfo {pages} {093020} (\bibinfo {year} {2011})}\BibitemShut {NoStop}%
\bibitem [{\citenamefont {Leymann}\ \emph {et~al.}(2015)\citenamefont {Leymann}, \citenamefont {Foerster}, \citenamefont {Jahnke}, \citenamefont {Wiersig},\ and\ \citenamefont {Gies}}]{Leymann2015Sub-Nanolasers}%
  \BibitemOpen
  \bibfield  {author} {\bibinfo {author} {\bibfnamefont {H.~A.~M.}\ \bibnamefont {Leymann}}, \bibinfo {author} {\bibfnamefont {A.}~\bibnamefont {Foerster}}, \bibinfo {author} {\bibfnamefont {F.}~\bibnamefont {Jahnke}}, \bibinfo {author} {\bibfnamefont {J.}~\bibnamefont {Wiersig}},\ and\ \bibinfo {author} {\bibfnamefont {C.}~\bibnamefont {Gies}},\ }\href {https://doi.org/10.1103/PhysRevApplied.4.044018} {\bibfield  {journal} {\bibinfo  {journal} {Phys. Rev. Appl.}\ }\textbf {\bibinfo {volume} {4}},\ \bibinfo {pages} {044018} (\bibinfo {year} {2015})}\BibitemShut {NoStop}%
\bibitem [{\citenamefont {Jahnke}\ \emph {et~al.}(2016)\citenamefont {Jahnke}, \citenamefont {Gies}, \citenamefont {A{\ss}mann}, \citenamefont {Bayer}, \citenamefont {Leymann}, \citenamefont {Foerster}, \citenamefont {Wiersig}, \citenamefont {Schneider}, \citenamefont {Kamp},\ and\ \citenamefont {H{\"o}fling}}]{Jahnke2016GiantNanolasers}%
  \BibitemOpen
  \bibfield  {author} {\bibinfo {author} {\bibfnamefont {F.}~\bibnamefont {Jahnke}}, \bibinfo {author} {\bibfnamefont {C.}~\bibnamefont {Gies}}, \bibinfo {author} {\bibfnamefont {M.}~\bibnamefont {A{\ss}mann}}, \bibinfo {author} {\bibfnamefont {M.}~\bibnamefont {Bayer}}, \bibinfo {author} {\bibfnamefont {H.~A.~M.}\ \bibnamefont {Leymann}}, \bibinfo {author} {\bibfnamefont {A.}~\bibnamefont {Foerster}}, \bibinfo {author} {\bibfnamefont {J.}~\bibnamefont {Wiersig}}, \bibinfo {author} {\bibfnamefont {C.}~\bibnamefont {Schneider}}, \bibinfo {author} {\bibfnamefont {M.}~\bibnamefont {Kamp}},\ and\ \bibinfo {author} {\bibfnamefont {S.}~\bibnamefont {H{\"o}fling}},\ }\href {https://doi.org/10.1038/ncomms11540} {\bibfield  {journal} {\bibinfo  {journal} {Nature Communications}\ }\textbf {\bibinfo {volume} {7}},\ \bibinfo {pages} {11540} (\bibinfo {year} {2016})}\BibitemShut {NoStop}%
\bibitem [{\citenamefont {Kreinberg}\ \emph {et~al.}(2017)\citenamefont {Kreinberg}, \citenamefont {Chow}, \citenamefont {Wolters}, \citenamefont {Schneider}, \citenamefont {Gies}, \citenamefont {Jahnke}, \citenamefont {H{\"o}fling}, \citenamefont {Kamp},\ and\ \citenamefont {Reitzenstein}}]{Kreinberg2017EmissionCoupling}%
  \BibitemOpen
  \bibfield  {author} {\bibinfo {author} {\bibfnamefont {S.}~\bibnamefont {Kreinberg}}, \bibinfo {author} {\bibfnamefont {W.~W.}\ \bibnamefont {Chow}}, \bibinfo {author} {\bibfnamefont {J.}~\bibnamefont {Wolters}}, \bibinfo {author} {\bibfnamefont {C.}~\bibnamefont {Schneider}}, \bibinfo {author} {\bibfnamefont {C.}~\bibnamefont {Gies}}, \bibinfo {author} {\bibfnamefont {F.}~\bibnamefont {Jahnke}}, \bibinfo {author} {\bibfnamefont {S.}~\bibnamefont {H{\"o}fling}}, \bibinfo {author} {\bibfnamefont {M.}~\bibnamefont {Kamp}},\ and\ \bibinfo {author} {\bibfnamefont {S.}~\bibnamefont {Reitzenstein}},\ }\href {https://doi.org/10.1038/lsa.2017.30} {\bibfield  {journal} {\bibinfo  {journal} {Light: Science \& Applications}\ }\textbf {\bibinfo {volume} {6}},\ \bibinfo {pages} {e17030} (\bibinfo {year} {2017})}\BibitemShut {NoStop}%
\bibitem [{\citenamefont {Zhang}\ \emph {et~al.}(2018)\citenamefont {Zhang}, \citenamefont {Zhang},\ and\ \citenamefont {M{\o}lmer}}]{Zhang2018MonteLasing}%
  \BibitemOpen
  \bibfield  {author} {\bibinfo {author} {\bibfnamefont {Y.}~\bibnamefont {Zhang}}, \bibinfo {author} {\bibfnamefont {Y.-X.}\ \bibnamefont {Zhang}},\ and\ \bibinfo {author} {\bibfnamefont {K.}~\bibnamefont {M{\o}lmer}},\ }\href {https://doi.org/10.1088/1367-2630/aaec36} {\bibfield  {journal} {\bibinfo  {journal} {New Journal of Physics}\ }\textbf {\bibinfo {volume} {20}},\ \bibinfo {pages} {112001} (\bibinfo {year} {2018})}\BibitemShut {NoStop}%
\bibitem [{\citenamefont {Andr{\'{e}}}\ \emph {et~al.}(2019)\citenamefont {Andr{\'{e}}}, \citenamefont {Protsenko}, \citenamefont {Uskov}, \citenamefont {M{\o}rk},\ and\ \citenamefont {Wubs}}]{Andre2019OnNano-LEDs}%
  \BibitemOpen
  \bibfield  {author} {\bibinfo {author} {\bibfnamefont {E.~C.}\ \bibnamefont {Andr{\'{e}}}}, \bibinfo {author} {\bibfnamefont {I.~E.}\ \bibnamefont {Protsenko}}, \bibinfo {author} {\bibfnamefont {A.~V.}\ \bibnamefont {Uskov}}, \bibinfo {author} {\bibfnamefont {J.}~\bibnamefont {M{\o}rk}},\ and\ \bibinfo {author} {\bibfnamefont {M.}~\bibnamefont {Wubs}},\ }\href {https://doi.org/10.1364/OL.44.001415} {\bibfield  {journal} {\bibinfo  {journal} {Optics Letters}\ }\textbf {\bibinfo {volume} {44}},\ \bibinfo {pages} {1415} (\bibinfo {year} {2019})}\BibitemShut {NoStop}%
\bibitem [{\citenamefont {Zhang}\ \emph {et~al.}(2021)\citenamefont {Zhang}, \citenamefont {Shan},\ and\ \citenamefont {M{\o}lmer}}]{Zhang2021UltranarrowStates}%
  \BibitemOpen
  \bibfield  {author} {\bibinfo {author} {\bibfnamefont {Y.}~\bibnamefont {Zhang}}, \bibinfo {author} {\bibfnamefont {C.}~\bibnamefont {Shan}},\ and\ \bibinfo {author} {\bibfnamefont {K.}~\bibnamefont {M{\o}lmer}},\ }\href {https://doi.org/10.1103/PhysRevLett.126.123602} {\bibfield  {journal} {\bibinfo  {journal} {Physical Review Letters}\ }\textbf {\bibinfo {volume} {126}},\ \bibinfo {pages} {123602} (\bibinfo {year} {2021})}\BibitemShut {NoStop}%
\bibitem [{\citenamefont {Protsenko}\ \emph {et~al.}(2021)\citenamefont {Protsenko}, \citenamefont {Uskov}, \citenamefont {Andr{\'{e}}}, \citenamefont {M{\o}rk},\ and\ \citenamefont {Wubs}}]{Protsenko2021QuantumNanolasers}%
  \BibitemOpen
  \bibfield  {author} {\bibinfo {author} {\bibfnamefont {I.~E.}\ \bibnamefont {Protsenko}}, \bibinfo {author} {\bibfnamefont {A.~V.}\ \bibnamefont {Uskov}}, \bibinfo {author} {\bibfnamefont {E.~C.}\ \bibnamefont {Andr{\'{e}}}}, \bibinfo {author} {\bibfnamefont {J.}~\bibnamefont {M{\o}rk}},\ and\ \bibinfo {author} {\bibfnamefont {M.}~\bibnamefont {Wubs}},\ }\href {https://doi.org/10.1088/1367-2630/abfd4c} {\bibfield  {journal} {\bibinfo  {journal} {New Journal of Physics}\ }\textbf {\bibinfo {volume} {23}},\ \bibinfo {pages} {063010} (\bibinfo {year} {2021})}\BibitemShut {NoStop}%
\bibitem [{\citenamefont {Vyshnevyy}(2022)}]{Vyshnevyy2022Gain-dependentNanolasers}%
  \BibitemOpen
  \bibfield  {author} {\bibinfo {author} {\bibfnamefont {A.~A.}\ \bibnamefont {Vyshnevyy}},\ }\href {https://doi.org/10.1103/PhysRevB.105.085116} {\bibfield  {journal} {\bibinfo  {journal} {Physical Review B}\ }\textbf {\bibinfo {volume} {105}},\ \bibinfo {pages} {085116} (\bibinfo {year} {2022})}\BibitemShut {NoStop}%
\bibitem [{\citenamefont {Meiser}\ \emph {et~al.}(2009)\citenamefont {Meiser}, \citenamefont {Ye}, \citenamefont {Carlson},\ and\ \citenamefont {Holland}}]{Meiser2009ProspectsLaser}%
  \BibitemOpen
  \bibfield  {author} {\bibinfo {author} {\bibfnamefont {D.}~\bibnamefont {Meiser}}, \bibinfo {author} {\bibfnamefont {J.}~\bibnamefont {Ye}}, \bibinfo {author} {\bibfnamefont {D.~R.}\ \bibnamefont {Carlson}},\ and\ \bibinfo {author} {\bibfnamefont {M.~J.}\ \bibnamefont {Holland}},\ }\href {https://doi.org/10.1103/PhysRevLett.102.163601} {\bibfield  {journal} {\bibinfo  {journal} {Physical Review Letters}\ }\textbf {\bibinfo {volume} {102}},\ \bibinfo {pages} {163601} (\bibinfo {year} {2009})}\BibitemShut {NoStop}%
\bibitem [{\citenamefont {Bohnet}\ \emph {et~al.}(2012)\citenamefont {Bohnet}, \citenamefont {Chen}, \citenamefont {Weiner}, \citenamefont {Meiser}, \citenamefont {Holland},\ and\ \citenamefont {Thompson}}]{Bohnet2012ALaser}%
  \BibitemOpen
  \bibfield  {author} {\bibinfo {author} {\bibfnamefont {J.~G.}\ \bibnamefont {Bohnet}}, \bibinfo {author} {\bibfnamefont {Z.}~\bibnamefont {Chen}}, \bibinfo {author} {\bibfnamefont {J.~M.}\ \bibnamefont {Weiner}}, \bibinfo {author} {\bibfnamefont {D.}~\bibnamefont {Meiser}}, \bibinfo {author} {\bibfnamefont {M.~J.}\ \bibnamefont {Holland}},\ and\ \bibinfo {author} {\bibfnamefont {J.~K.}\ \bibnamefont {Thompson}},\ }\href {https://doi.org/10.1038/nature10920} {\bibfield  {journal} {\bibinfo  {journal} {Nature}\ }\textbf {\bibinfo {volume} {484}},\ \bibinfo {pages} {78} (\bibinfo {year} {2012})}\BibitemShut {NoStop}%
\bibitem [{\citenamefont {Norcia}\ \emph {et~al.}(2016)\citenamefont {Norcia}, \citenamefont {Winchester}, \citenamefont {Cline},\ and\ \citenamefont {Thompson}}]{Norcia2016SuperradianceLaser}%
  \BibitemOpen
  \bibfield  {author} {\bibinfo {author} {\bibfnamefont {M.~A.}\ \bibnamefont {Norcia}}, \bibinfo {author} {\bibfnamefont {M.~N.}\ \bibnamefont {Winchester}}, \bibinfo {author} {\bibfnamefont {J.~R.~K.}\ \bibnamefont {Cline}},\ and\ \bibinfo {author} {\bibfnamefont {J.~K.}\ \bibnamefont {Thompson}},\ }\href {https://doi.org/10.1126/sciadv.1601231} {\bibfield  {journal} {\bibinfo  {journal} {Science Advances}\ }\textbf {\bibinfo {volume} {2}},\ \bibinfo {pages} {e1601231} (\bibinfo {year} {2016})}\BibitemShut {NoStop}%
\bibitem [{\citenamefont {Debnath}\ \emph {et~al.}(2018)\citenamefont {Debnath}, \citenamefont {Zhang},\ and\ \citenamefont {M{\o}lmer}}]{Debnath2018LasingRegime}%
  \BibitemOpen
  \bibfield  {author} {\bibinfo {author} {\bibfnamefont {K.}~\bibnamefont {Debnath}}, \bibinfo {author} {\bibfnamefont {Y.}~\bibnamefont {Zhang}},\ and\ \bibinfo {author} {\bibfnamefont {K.}~\bibnamefont {M{\o}lmer}},\ }\href {https://doi.org/10.1103/PhysRevA.98.063837} {\bibfield  {journal} {\bibinfo  {journal} {Physical Review A}\ }\textbf {\bibinfo {volume} {98}},\ \bibinfo {pages} {063837} (\bibinfo {year} {2018})}\BibitemShut {NoStop}%
\bibitem [{\citenamefont {Chase}\ and\ \citenamefont {Geremia}(2008)}]{Chase2008CollectiveParticles}%
  \BibitemOpen
  \bibfield  {author} {\bibinfo {author} {\bibfnamefont {B.~A.}\ \bibnamefont {Chase}}\ and\ \bibinfo {author} {\bibfnamefont {J.~M.}\ \bibnamefont {Geremia}},\ }\href {https://doi.org/10.1103/PhysRevA.78.052101} {\bibfield  {journal} {\bibinfo  {journal} {Physical Review A}\ }\textbf {\bibinfo {volume} {78}},\ \bibinfo {pages} {052101} (\bibinfo {year} {2008})}\BibitemShut {NoStop}%
\bibitem [{\citenamefont {Baragiola}\ \emph {et~al.}(2010)\citenamefont {Baragiola}, \citenamefont {Chase},\ and\ \citenamefont {Geremia}}]{Baragiola2010CollectiveSystems}%
  \BibitemOpen
  \bibfield  {author} {\bibinfo {author} {\bibfnamefont {B.~Q.}\ \bibnamefont {Baragiola}}, \bibinfo {author} {\bibfnamefont {B.~A.}\ \bibnamefont {Chase}},\ and\ \bibinfo {author} {\bibfnamefont {J.~M.}\ \bibnamefont {Geremia}},\ }\href {https://doi.org/10.1103/PhysRevA.81.032104} {\bibfield  {journal} {\bibinfo  {journal} {Physical Review A}\ }\textbf {\bibinfo {volume} {81}},\ \bibinfo {pages} {032104} (\bibinfo {year} {2010})}\BibitemShut {NoStop}%
\bibitem [{\citenamefont {Gegg}\ and\ \citenamefont {Richter}(2016)}]{Gegg2016EfficientSystems}%
  \BibitemOpen
  \bibfield  {author} {\bibinfo {author} {\bibfnamefont {M.}~\bibnamefont {Gegg}}\ and\ \bibinfo {author} {\bibfnamefont {M.}~\bibnamefont {Richter}},\ }\href {https://doi.org/10.1088/1367-2630/18/4/043037} {\bibfield  {journal} {\bibinfo  {journal} {New Journal of Physics}\ }\textbf {\bibinfo {volume} {18}},\ \bibinfo {pages} {043037} (\bibinfo {year} {2016})}\BibitemShut {NoStop}%
\bibitem [{\citenamefont {Kirton}\ and\ \citenamefont {Keeling}(2017)}]{Kirton2017SuppressingDecay}%
  \BibitemOpen
  \bibfield  {author} {\bibinfo {author} {\bibfnamefont {P.}~\bibnamefont {Kirton}}\ and\ \bibinfo {author} {\bibfnamefont {J.}~\bibnamefont {Keeling}},\ }\href {https://doi.org/10.1103/PhysRevLett.118.123602} {\bibfield  {journal} {\bibinfo  {journal} {Physical Review Letters}\ }\textbf {\bibinfo {volume} {118}},\ \bibinfo {pages} {123602} (\bibinfo {year} {2017})}\BibitemShut {NoStop}%
\bibitem [{\citenamefont {Shammah}\ \emph {et~al.}(2018)\citenamefont {Shammah}, \citenamefont {Ahmed}, \citenamefont {Lambert}, \citenamefont {De~Liberato},\ and\ \citenamefont {Nori}}]{Shammah2018OpenInvariance}%
  \BibitemOpen
  \bibfield  {author} {\bibinfo {author} {\bibfnamefont {N.}~\bibnamefont {Shammah}}, \bibinfo {author} {\bibfnamefont {S.}~\bibnamefont {Ahmed}}, \bibinfo {author} {\bibfnamefont {N.}~\bibnamefont {Lambert}}, \bibinfo {author} {\bibfnamefont {S.}~\bibnamefont {De~Liberato}},\ and\ \bibinfo {author} {\bibfnamefont {F.}~\bibnamefont {Nori}},\ }\href {https://doi.org/10.1103/PhysRevA.98.063815} {\bibfield  {journal} {\bibinfo  {journal} {Physical Review A}\ }\textbf {\bibinfo {volume} {98}},\ \bibinfo {pages} {063815} (\bibinfo {year} {2018})}\BibitemShut {NoStop}%
\bibitem [{\citenamefont {Gies}\ \emph {et~al.}(2007)\citenamefont {Gies}, \citenamefont {Wiersig}, \citenamefont {Lorke},\ and\ \citenamefont {Jahnke}}]{Gies2007SemiconductorLasers}%
  \BibitemOpen
  \bibfield  {author} {\bibinfo {author} {\bibfnamefont {C.}~\bibnamefont {Gies}}, \bibinfo {author} {\bibfnamefont {J.}~\bibnamefont {Wiersig}}, \bibinfo {author} {\bibfnamefont {M.}~\bibnamefont {Lorke}},\ and\ \bibinfo {author} {\bibfnamefont {F.}~\bibnamefont {Jahnke}},\ }\href {https://doi.org/10.1103/PhysRevA.75.013803} {\bibfield  {journal} {\bibinfo  {journal} {Physical Review A}\ }\textbf {\bibinfo {volume} {75}},\ \bibinfo {pages} {013803} (\bibinfo {year} {2007})}\BibitemShut {NoStop}%
\bibitem [{\citenamefont {Plankensteiner}\ \emph {et~al.}(2022)\citenamefont {Plankensteiner}, \citenamefont {Hotter},\ and\ \citenamefont {Ritsch}}]{Plankensteiner2022quantumcumulantsjl}%
  \BibitemOpen
  \bibfield  {author} {\bibinfo {author} {\bibfnamefont {D.}~\bibnamefont {Plankensteiner}}, \bibinfo {author} {\bibfnamefont {C.}~\bibnamefont {Hotter}},\ and\ \bibinfo {author} {\bibfnamefont {H.}~\bibnamefont {Ritsch}},\ }\href {https://doi.org/10.22331/q-2022-01-04-617} {\bibfield  {journal} {\bibinfo  {journal} {Quantum}\ }\textbf {\bibinfo {volume} {6}},\ \bibinfo {pages} {617} (\bibinfo {year} {2022})}\BibitemShut {NoStop}%
\bibitem [{\citenamefont {Coldren}\ \emph {et~al.}(2012)\citenamefont {Coldren}, \citenamefont {Corzine},\ and\ \citenamefont {Ma{\v{s}}anovi{\'c}}}]{Coldren1997}%
  \BibitemOpen
  \bibfield  {author} {\bibinfo {author} {\bibfnamefont {L.~A.}\ \bibnamefont {Coldren}}, \bibinfo {author} {\bibfnamefont {S.~W.}\ \bibnamefont {Corzine}},\ and\ \bibinfo {author} {\bibfnamefont {M.~L.}\ \bibnamefont {Ma{\v{s}}anovi{\'c}}},\ }\href@noop {} {\emph {\bibinfo {title} {Diode Lasers and Photonic Integrated Circuits}}},\ \bibinfo {edition} {2nd}\ ed.\ (\bibinfo  {publisher} {Wiley},\ \bibinfo {year} {2012})\BibitemShut {NoStop}%
\bibitem [{\citenamefont {Bundgaard-Nielsen}\ \emph {et~al.}(2023)\citenamefont {Bundgaard-Nielsen}, \citenamefont {Denning}, \citenamefont {Saldutti},\ and\ \citenamefont {M{\o}rk}}]{Bundgaard-Nielsen2023StochasticNanolaser}%
  \BibitemOpen
  \bibfield  {author} {\bibinfo {author} {\bibfnamefont {M.}~\bibnamefont {Bundgaard-Nielsen}}, \bibinfo {author} {\bibfnamefont {E.~V.}\ \bibnamefont {Denning}}, \bibinfo {author} {\bibfnamefont {M.}~\bibnamefont {Saldutti}},\ and\ \bibinfo {author} {\bibfnamefont {J.}~\bibnamefont {M{\o}rk}},\ }\href {https://doi.org/10.1103/PhysRevLett.130.253801} {\bibfield  {journal} {\bibinfo  {journal} {Physical Review Letters}\ }\textbf {\bibinfo {volume} {130}},\ \bibinfo {pages} {253801} (\bibinfo {year} {2023})}\BibitemShut {NoStop}%
\bibitem [{\citenamefont {M{\o}rk}\ and\ \citenamefont {Lippi}(2018)}]{Mork2018RateEmitters}%
  \BibitemOpen
  \bibfield  {author} {\bibinfo {author} {\bibfnamefont {J.}~\bibnamefont {M{\o}rk}}\ and\ \bibinfo {author} {\bibfnamefont {G.~L.}\ \bibnamefont {Lippi}},\ }\href {https://doi.org/10.1063/1.5022958} {\bibfield  {journal} {\bibinfo  {journal} {Applied Physics Letters}\ }\textbf {\bibinfo {volume} {112}},\ \bibinfo {pages} {141103} (\bibinfo {year} {2018})}\BibitemShut {NoStop}%
\bibitem [{\citenamefont {Puccioni}\ and\ \citenamefont {Lippi}(2015)}]{Puccioni2015StochasticLasing}%
  \BibitemOpen
  \bibfield  {author} {\bibinfo {author} {\bibfnamefont {G.~P.}\ \bibnamefont {Puccioni}}\ and\ \bibinfo {author} {\bibfnamefont {G.~L.}\ \bibnamefont {Lippi}},\ }\href {https://doi.org/10.1364/OE.23.002369} {\bibfield  {journal} {\bibinfo  {journal} {Optics Express}\ }\textbf {\bibinfo {volume} {23}},\ \bibinfo {pages} {2369} (\bibinfo {year} {2015})}\BibitemShut {NoStop}%
\bibitem [{\citenamefont {Andr{\'{e}}}\ \emph {et~al.}(2020)\citenamefont {Andr{\'{e}}}, \citenamefont {M{\o}rk},\ and\ \citenamefont {Wubs}}]{Andre2020}%
  \BibitemOpen
  \bibfield  {author} {\bibinfo {author} {\bibfnamefont {E.~C.}\ \bibnamefont {Andr{\'{e}}}}, \bibinfo {author} {\bibfnamefont {J.}~\bibnamefont {M{\o}rk}},\ and\ \bibinfo {author} {\bibfnamefont {M.}~\bibnamefont {Wubs}},\ }\href {https://doi.org/10.1364/OE.405979} {\bibfield  {journal} {\bibinfo  {journal} {Optics Express}\ }\textbf {\bibinfo {volume} {28}},\ \bibinfo {pages} {32632} (\bibinfo {year} {2020})}\BibitemShut {NoStop}%
\bibitem [{\citenamefont {Bundgaard-Nielsen}\ \emph {et~al.}(2026)\citenamefont {Bundgaard-Nielsen}, \citenamefont {Lippi},\ and\ \citenamefont {M{\o}rk}}]{Bundgaard-Nielsen2026ModelingChains}%
  \BibitemOpen
  \bibfield  {author} {\bibinfo {author} {\bibfnamefont {M.}~\bibnamefont {Bundgaard-Nielsen}}, \bibinfo {author} {\bibfnamefont {G.~L.}\ \bibnamefont {Lippi}},\ and\ \bibinfo {author} {\bibfnamefont {J.}~\bibnamefont {M{\o}rk}},\ }\href {https://doi.org/10.1103/f6f5-8dv8} {\bibfield  {journal} {\bibinfo  {journal} {Physical Review A}\ }\textbf {\bibinfo {volume} {113}},\ \bibinfo {pages} {043704} (\bibinfo {year} {2026})}\BibitemShut {NoStop}%
\bibitem [{\citenamefont {Bundgaard-Nielsen}\ \emph {et~al.}(2025)\citenamefont {Bundgaard-Nielsen}, \citenamefont {Saldutti}, \citenamefont {G{\o}tzsche}, \citenamefont {Grovn},\ and\ \citenamefont {M{\o}rk}}]{Bundgaard-Nielsen2025SimpleNanolasers}%
  \BibitemOpen
  \bibfield  {author} {\bibinfo {author} {\bibfnamefont {M.}~\bibnamefont {Bundgaard-Nielsen}}, \bibinfo {author} {\bibfnamefont {M.}~\bibnamefont {Saldutti}}, \bibinfo {author} {\bibfnamefont {B.~F.}\ \bibnamefont {G{\o}tzsche}}, \bibinfo {author} {\bibfnamefont {E.}~\bibnamefont {Grovn}},\ and\ \bibinfo {author} {\bibfnamefont {J.}~\bibnamefont {M{\o}rk}},\ }\href {https://doi.org/10.1103/PhysRevLett.134.213804} {\bibfield  {journal} {\bibinfo  {journal} {Physical Review Letters}\ }\textbf {\bibinfo {volume} {134}},\ \bibinfo {pages} {213804} (\bibinfo {year} {2025})}\BibitemShut {NoStop}%
\bibitem [{\citenamefont {Bundgaard-Nielsen}(2026)}]{Bundgaard-Nielsen2026SuperradiantLMCCode}%
  \BibitemOpen
  \bibfield  {author} {\bibinfo {author} {\bibfnamefont {M.}~\bibnamefont {Bundgaard-Nielsen}},\ }\href@noop {} {\bibinfo {title} {{SuperradiantLMC}: {D}icke laser {M}arkov chains and the polarization-order hierarchy of the laser master equation}},\ \bibinfo {howpublished} {\url{https://github.com/mabuni1998/superradiant-lmc}} (\bibinfo {year} {2026})\BibitemShut {NoStop}%
\bibitem [{\citenamefont {Tavis}\ and\ \citenamefont {Cummings}(1968)}]{TavisExactHamiltonian1968}%
  \BibitemOpen
  \bibfield  {author} {\bibinfo {author} {\bibfnamefont {M.}~\bibnamefont {Tavis}}\ and\ \bibinfo {author} {\bibfnamefont {F.~W.}\ \bibnamefont {Cummings}},\ }\href {https://doi.org/10.1103/PhysRev.170.379} {\bibfield  {journal} {\bibinfo  {journal} {Phys. Rev.}\ }\textbf {\bibinfo {volume} {170}},\ \bibinfo {pages} {379} (\bibinfo {year} {1968})}\BibitemShut {NoStop}%
\bibitem [{\citenamefont {Moelbjerg}\ \emph {et~al.}(2013)\citenamefont {Moelbjerg}, \citenamefont {Kaer}, \citenamefont {Lorke}, \citenamefont {Tromborg},\ and\ \citenamefont {Mork}}]{Moelbjerg2013DynamicalEmitters}%
  \BibitemOpen
  \bibfield  {author} {\bibinfo {author} {\bibfnamefont {A.}~\bibnamefont {Moelbjerg}}, \bibinfo {author} {\bibfnamefont {P.}~\bibnamefont {Kaer}}, \bibinfo {author} {\bibfnamefont {M.}~\bibnamefont {Lorke}}, \bibinfo {author} {\bibfnamefont {B.}~\bibnamefont {Tromborg}},\ and\ \bibinfo {author} {\bibfnamefont {J.}~\bibnamefont {Mork}},\ }\href {https://doi.org/10.1109/JQE.2013.2282464} {\bibfield  {journal} {\bibinfo  {journal} {IEEE Journal of Quantum Electronics}\ }\textbf {\bibinfo {volume} {49}},\ \bibinfo {pages} {945} (\bibinfo {year} {2013})}\BibitemShut {NoStop}%
\bibitem [{\citenamefont {Mu}\ and\ \citenamefont {Savage}(1992)}]{Mu1992One-atomLasers}%
  \BibitemOpen
  \bibfield  {author} {\bibinfo {author} {\bibfnamefont {Y.}~\bibnamefont {Mu}}\ and\ \bibinfo {author} {\bibfnamefont {C.~M.}\ \bibnamefont {Savage}},\ }\href {https://doi.org/10.1103/PhysRevA.46.5944} {\bibfield  {journal} {\bibinfo  {journal} {Physical Review A}\ }\textbf {\bibinfo {volume} {46}},\ \bibinfo {pages} {5944} (\bibinfo {year} {1992})}\BibitemShut {NoStop}%
\bibitem [{\citenamefont {L{\"o}ffler}\ \emph {et~al.}(1997)\citenamefont {L{\"o}ffler}, \citenamefont {Meyer},\ and\ \citenamefont {Walther}}]{Loffler1997SpectralLaser}%
  \BibitemOpen
  \bibfield  {author} {\bibinfo {author} {\bibfnamefont {M.}~\bibnamefont {L{\"o}ffler}}, \bibinfo {author} {\bibfnamefont {G.~M.}\ \bibnamefont {Meyer}},\ and\ \bibinfo {author} {\bibfnamefont {H.}~\bibnamefont {Walther}},\ }\href {https://doi.org/10.1103/PhysRevA.55.3923} {\bibfield  {journal} {\bibinfo  {journal} {Physical Review A}\ }\textbf {\bibinfo {volume} {55}},\ \bibinfo {pages} {3923} (\bibinfo {year} {1997})}\BibitemShut {NoStop}%
\bibitem [{\citenamefont {Nomura}\ \emph {et~al.}(2010)\citenamefont {Nomura}, \citenamefont {Kumagai}, \citenamefont {Iwamoto}, \citenamefont {Ota},\ and\ \citenamefont {Arakawa}}]{Nomura2010LaserSystem}%
  \BibitemOpen
  \bibfield  {author} {\bibinfo {author} {\bibfnamefont {M.}~\bibnamefont {Nomura}}, \bibinfo {author} {\bibfnamefont {N.}~\bibnamefont {Kumagai}}, \bibinfo {author} {\bibfnamefont {S.}~\bibnamefont {Iwamoto}}, \bibinfo {author} {\bibfnamefont {Y.}~\bibnamefont {Ota}},\ and\ \bibinfo {author} {\bibfnamefont {Y.}~\bibnamefont {Arakawa}},\ }\href {https://doi.org/10.1038/nphys1518} {\bibfield  {journal} {\bibinfo  {journal} {Nature Physics}\ }\textbf {\bibinfo {volume} {6}},\ \bibinfo {pages} {279} (\bibinfo {year} {2010})}\BibitemShut {NoStop}%
\bibitem [{\citenamefont {Johansson}\ \emph {et~al.}(2012)\citenamefont {Johansson}, \citenamefont {Nation},\ and\ \citenamefont {Nori}}]{Johansson2012QuTiP}%
  \BibitemOpen
  \bibfield  {author} {\bibinfo {author} {\bibfnamefont {J.~R.}\ \bibnamefont {Johansson}}, \bibinfo {author} {\bibfnamefont {P.~D.}\ \bibnamefont {Nation}},\ and\ \bibinfo {author} {\bibfnamefont {F.}~\bibnamefont {Nori}},\ }\href {https://doi.org/10.1016/j.cpc.2012.02.021} {\bibfield  {journal} {\bibinfo  {journal} {Computer Physics Communications}\ }\textbf {\bibinfo {volume} {183}},\ \bibinfo {pages} {1760} (\bibinfo {year} {2012})}\BibitemShut {NoStop}%
\bibitem [{\citenamefont {Roy-Choudhury}\ \emph {et~al.}(2009)\citenamefont {Roy-Choudhury}, \citenamefont {Haas},\ and\ \citenamefont {Levi}}]{Roy-Choudhury2009QuantumLasers}%
  \BibitemOpen
  \bibfield  {author} {\bibinfo {author} {\bibfnamefont {K.}~\bibnamefont {Roy-Choudhury}}, \bibinfo {author} {\bibfnamefont {S.}~\bibnamefont {Haas}},\ and\ \bibinfo {author} {\bibfnamefont {A.~F.~J.}\ \bibnamefont {Levi}},\ }\href {https://doi.org/10.1103/PhysRevLett.102.053902} {\bibfield  {journal} {\bibinfo  {journal} {Physical Review Letters}\ }\textbf {\bibinfo {volume} {102}},\ \bibinfo {pages} {053902} (\bibinfo {year} {2009})}\BibitemShut {NoStop}%
\bibitem [{\citenamefont {Roy-Choudhury}\ and\ \citenamefont {Levi}(2010)}]{Roy-Choudhury2010QuantumDiodes}%
  \BibitemOpen
  \bibfield  {author} {\bibinfo {author} {\bibfnamefont {K.}~\bibnamefont {Roy-Choudhury}}\ and\ \bibinfo {author} {\bibfnamefont {A.~F.~J.}\ \bibnamefont {Levi}},\ }\href {https://doi.org/10.1103/PhysRevA.81.013827} {\bibfield  {journal} {\bibinfo  {journal} {Phys. Rev. A}\ }\textbf {\bibinfo {volume} {81}},\ \bibinfo {pages} {013827} (\bibinfo {year} {2010})}\BibitemShut {NoStop}%
\bibitem [{\citenamefont {Lebreton}\ \emph {et~al.}(2013)\citenamefont {Lebreton}, \citenamefont {Abram}, \citenamefont {Takemura}, \citenamefont {Kuwata-Gonokami}, \citenamefont {Robert-Philip},\ and\ \citenamefont {Beveratos}}]{Lebreton2013StochasticallyNanolasers}%
  \BibitemOpen
  \bibfield  {author} {\bibinfo {author} {\bibfnamefont {A.}~\bibnamefont {Lebreton}}, \bibinfo {author} {\bibfnamefont {I.}~\bibnamefont {Abram}}, \bibinfo {author} {\bibfnamefont {N.}~\bibnamefont {Takemura}}, \bibinfo {author} {\bibfnamefont {M.}~\bibnamefont {Kuwata-Gonokami}}, \bibinfo {author} {\bibfnamefont {I.}~\bibnamefont {Robert-Philip}},\ and\ \bibinfo {author} {\bibfnamefont {A.}~\bibnamefont {Beveratos}},\ }\href {https://doi.org/10.1088/1367-2630/15/3/033039} {\bibfield  {journal} {\bibinfo  {journal} {New Journal of Physics}\ }\textbf {\bibinfo {volume} {15}},\ \bibinfo {pages} {033039} (\bibinfo {year} {2013})}\BibitemShut {NoStop}%
\bibitem [{\citenamefont {Takemura}\ \emph {et~al.}(2019)\citenamefont {Takemura}, \citenamefont {Takiguchi}, \citenamefont {Kuramochi}, \citenamefont {Shinya}, \citenamefont {Sato}, \citenamefont {Takeda}, \citenamefont {Matsuo},\ and\ \citenamefont {Notomi}}]{Takemura2019LasingLasers}%
  \BibitemOpen
  \bibfield  {author} {\bibinfo {author} {\bibfnamefont {N.}~\bibnamefont {Takemura}}, \bibinfo {author} {\bibfnamefont {M.}~\bibnamefont {Takiguchi}}, \bibinfo {author} {\bibfnamefont {E.}~\bibnamefont {Kuramochi}}, \bibinfo {author} {\bibfnamefont {A.}~\bibnamefont {Shinya}}, \bibinfo {author} {\bibfnamefont {T.}~\bibnamefont {Sato}}, \bibinfo {author} {\bibfnamefont {K.}~\bibnamefont {Takeda}}, \bibinfo {author} {\bibfnamefont {S.}~\bibnamefont {Matsuo}},\ and\ \bibinfo {author} {\bibfnamefont {M.}~\bibnamefont {Notomi}},\ }\href {https://doi.org/10.1103/PhysRevA.99.053820} {\bibfield  {journal} {\bibinfo  {journal} {Physical Review A}\ }\textbf {\bibinfo {volume} {99}},\ \bibinfo {pages} {053820} (\bibinfo {year} {2019})}\BibitemShut {NoStop}%
\bibitem [{\citenamefont {Lippi}(2021)}]{Lippi2021AmplifiedNanolasers}%
  \BibitemOpen
  \bibfield  {author} {\bibinfo {author} {\bibfnamefont {G.~L.}\ \bibnamefont {Lippi}},\ }\href {https://doi.org/10.3390/atoms9010006} {\bibfield  {journal} {\bibinfo  {journal} {Atoms}\ }\textbf {\bibinfo {volume} {9}},\ \bibinfo {pages} {6} (\bibinfo {year} {2021})}\BibitemShut {NoStop}%
\bibitem [{\citenamefont {Puccioni}\ \emph {et~al.}(2024)\citenamefont {Puccioni}, \citenamefont {Wang},\ and\ \citenamefont {Lippi}}]{Puccioni2024StatisticsNanolasers}%
  \BibitemOpen
  \bibfield  {author} {\bibinfo {author} {\bibfnamefont {G.}~\bibnamefont {Puccioni}}, \bibinfo {author} {\bibfnamefont {T.}~\bibnamefont {Wang}},\ and\ \bibinfo {author} {\bibfnamefont {G.}~\bibnamefont {Lippi}},\ }in\ \href {https://doi.org/10.1117/12.3023714} {\emph {\bibinfo {booktitle} {Semiconductor Lasers and Laser Dynamics XI}}},\ Vol.\ \bibinfo {volume} {13002},\ \bibinfo {editor} {edited by\ \bibinfo {editor} {\bibfnamefont {M.}~\bibnamefont {Sciamanna}}, \bibinfo {editor} {\bibfnamefont {F.-Y.}\ \bibnamefont {Lin}},\ and\ \bibinfo {editor} {\bibfnamefont {J.}~\bibnamefont {M{\o}rk}}},\ \bibinfo {organization} {International Society for Optics and Photonics}\ (\bibinfo  {publisher} {SPIE},\ \bibinfo {year} {2024})\ p.\ \bibinfo {pages} {1300207}\BibitemShut {NoStop}%
\bibitem [{\citenamefont {Strauf}\ and\ \citenamefont {Jahnke}(2011)}]{Strauf2011SingleNanolaser}%
  \BibitemOpen
  \bibfield  {author} {\bibinfo {author} {\bibfnamefont {S.}~\bibnamefont {Strauf}}\ and\ \bibinfo {author} {\bibfnamefont {F.}~\bibnamefont {Jahnke}},\ }\href {https://doi.org/10.1002/lpor.201000039} {\bibfield  {journal} {\bibinfo  {journal} {Laser \& Photonics Reviews}\ }\textbf {\bibinfo {volume} {5}},\ \bibinfo {pages} {607} (\bibinfo {year} {2011})}\BibitemShut {NoStop}%
\bibitem [{\citenamefont {Gillespie}(1976)}]{Gillespie1976AReactions}%
  \BibitemOpen
  \bibfield  {author} {\bibinfo {author} {\bibfnamefont {D.~T.}\ \bibnamefont {Gillespie}},\ }\href {https://doi.org/10.1016/0021-9991(76)90041-3} {\bibfield  {journal} {\bibinfo  {journal} {Journal of Computational Physics}\ }\textbf {\bibinfo {volume} {22}},\ \bibinfo {pages} {403} (\bibinfo {year} {1976})}\BibitemShut {NoStop}%
\bibitem [{\citenamefont {Gillespie}(1992)}]{Gillespie1992AEquation}%
  \BibitemOpen
  \bibfield  {author} {\bibinfo {author} {\bibfnamefont {D.~T.}\ \bibnamefont {Gillespie}},\ }\href {https://doi.org/10.1016/0378-4371(92)90283-V} {\bibfield  {journal} {\bibinfo  {journal} {Physica A: Statistical Mechanics and its Applications}\ }\textbf {\bibinfo {volume} {188}},\ \bibinfo {pages} {404} (\bibinfo {year} {1992})}\BibitemShut {NoStop}%
\bibitem [{\citenamefont {Gillespie}(2007)}]{Gillespie2007StochasticKinetics}%
  \BibitemOpen
  \bibfield  {author} {\bibinfo {author} {\bibfnamefont {D.~T.}\ \bibnamefont {Gillespie}},\ }\href {https://doi.org/10.1146/annurev.physchem.58.032806.104637} {\bibfield  {journal} {\bibinfo  {journal} {Annual Review of Physical Chemistry}\ }\textbf {\bibinfo {volume} {58}},\ \bibinfo {pages} {35} (\bibinfo {year} {2007})}\BibitemShut {NoStop}%
\bibitem [{\citenamefont {M{\o}lmer}\ \emph {et~al.}(1993)\citenamefont {M{\o}lmer}, \citenamefont {Castin},\ and\ \citenamefont {Dalibard}}]{Mlmer1993MonteOptics}%
  \BibitemOpen
  \bibfield  {author} {\bibinfo {author} {\bibfnamefont {K.}~\bibnamefont {M{\o}lmer}}, \bibinfo {author} {\bibfnamefont {Y.}~\bibnamefont {Castin}},\ and\ \bibinfo {author} {\bibfnamefont {J.}~\bibnamefont {Dalibard}},\ }\href {https://doi.org/10.1364/josab.10.000524} {\bibfield  {journal} {\bibinfo  {journal} {Journal of the Optical Society of America B}\ }\textbf {\bibinfo {volume} {10}},\ \bibinfo {pages} {524} (\bibinfo {year} {1993})}\BibitemShut {NoStop}%
\bibitem [{\citenamefont {Kuppens}\ \emph {et~al.}(1994)\citenamefont {Kuppens}, \citenamefont {van Exter},\ and\ \citenamefont {Woerdman}}]{Kuppens1994Quantum-limitedLaser}%
  \BibitemOpen
  \bibfield  {author} {\bibinfo {author} {\bibfnamefont {S.~J.~M.}\ \bibnamefont {Kuppens}}, \bibinfo {author} {\bibfnamefont {M.~P.}\ \bibnamefont {van Exter}},\ and\ \bibinfo {author} {\bibfnamefont {J.~P.}\ \bibnamefont {Woerdman}},\ }\href {https://doi.org/10.1103/PhysRevLett.72.3815} {\bibfield  {journal} {\bibinfo  {journal} {Physical Review Letters}\ }\textbf {\bibinfo {volume} {72}},\ \bibinfo {pages} {3815} (\bibinfo {year} {1994})}\BibitemShut {NoStop}%
\bibitem [{\citenamefont {Debnath}\ \emph {et~al.}(2019)\citenamefont {Debnath}, \citenamefont {Zhang},\ and\ \citenamefont {M{\o}lmer}}]{Debnath2019CollectiveLinewidth}%
  \BibitemOpen
  \bibfield  {author} {\bibinfo {author} {\bibfnamefont {K.}~\bibnamefont {Debnath}}, \bibinfo {author} {\bibfnamefont {Y.}~\bibnamefont {Zhang}},\ and\ \bibinfo {author} {\bibfnamefont {K.}~\bibnamefont {M{\o}lmer}},\ }\href {https://doi.org/10.1103/PhysRevA.100.053821} {\bibfield  {journal} {\bibinfo  {journal} {Physical Review A}\ }\textbf {\bibinfo {volume} {100}},\ \bibinfo {pages} {053821} (\bibinfo {year} {2019})}\BibitemShut {NoStop}%
\bibitem [{\citenamefont {Bychek}\ \emph {et~al.}(2021)\citenamefont {Bychek}, \citenamefont {Hotter}, \citenamefont {Plankensteiner},\ and\ \citenamefont {Ritsch}}]{Bychek_Hotter_Plankensteiner_Ritsch_2021}%
  \BibitemOpen
  \bibfield  {author} {\bibinfo {author} {\bibfnamefont {A.}~\bibnamefont {Bychek}}, \bibinfo {author} {\bibfnamefont {C.}~\bibnamefont {Hotter}}, \bibinfo {author} {\bibfnamefont {D.}~\bibnamefont {Plankensteiner}},\ and\ \bibinfo {author} {\bibfnamefont {H.}~\bibnamefont {Ritsch}},\ }\bibfield  {journal} {\bibinfo  {journal} {Open Research Europe}\ }\href {https://doi.org/10.12688/openreseurope.13781.2} {10.12688/openreseurope.13781.2} (\bibinfo {year} {2021})\BibitemShut {NoStop}%
\bibitem [{\citenamefont {Albert}\ and\ \citenamefont {Jiang}(2014)}]{Albert2014SymmetriesEquations}%
  \BibitemOpen
  \bibfield  {author} {\bibinfo {author} {\bibfnamefont {V.~V.}\ \bibnamefont {Albert}}\ and\ \bibinfo {author} {\bibfnamefont {L.}~\bibnamefont {Jiang}},\ }\href {https://doi.org/10.1103/PhysRevA.89.022118} {\bibfield  {journal} {\bibinfo  {journal} {Physical Review A}\ }\textbf {\bibinfo {volume} {89}},\ \bibinfo {pages} {022118} (\bibinfo {year} {2014})}\BibitemShut {NoStop}%
\bibitem [{\citenamefont {Gillespie}(2001)}]{Gillespie2001ApproximateSystems}%
  \BibitemOpen
  \bibfield  {author} {\bibinfo {author} {\bibfnamefont {D.~T.}\ \bibnamefont {Gillespie}},\ }\href {https://doi.org/10.1063/1.1378322} {\bibfield  {journal} {\bibinfo  {journal} {The Journal of Chemical Physics}\ }\textbf {\bibinfo {volume} {115}},\ \bibinfo {pages} {1716} (\bibinfo {year} {2001})}\BibitemShut {NoStop}%
\end{thebibliography}
\end{document}